\documentclass[sigconf]{acmart}

\AtBeginDocument{%
  \providecommand\BibTeX{{%
    \normalfont B\kern-0.5em{\scshape i\kern-0.25em b}\kern-0.8em\TeX}}}

\setcopyright{cc}
\copyrightyear{2022}
\acmYear{2022}
\acmDOI{10.1145/3545948.3545968}

\acmConference[RAID 2022]{25th International Symposium on Research in Attacks, Intrusions and Defenses}{October 26--28, 2022}{Limassol, Cyprus}
\acmBooktitle{25th International Symposium on Research in Attacks, Intrusions and Defenses (RAID 2022), October 26--28, 2022, Limassol, Cyprus}
\acmISBN{978-1-4503-9704-9/22/10}

\usepackage[nolist]{acronym} %
\usepackage{amsmath}
\usepackage[inline]{enumitem} %
\usepackage[normalem]{ulem} %
\usepackage{subcaption}
\usepackage{tikz}
\usepackage{xspace}

\usepackage{lipsum}

\newcommand{\etal}{et~al.\xspace}

\usepackage{adjustbox}
\usepackage{booktabs}
\usepackage{wasysym}
\usepackage{tabularx} %
\usepackage{multirow}
\usepackage{arydshln} %

\newcommand\Tstrut{\rule{0pt}{6pt}}         %

\newcommand{\bad}{$\Circle$}
\newcommand{\ok}{$\LEFTcircle$}
\newcommand{\good}{$\CIRCLE$}

\definecolor{gray1}{gray}{.95}
\definecolor{gray2}{gray}{.85}
\definecolor{gray3}{gray}{.50}
\definecolor{lightblue}{HTML}{2596be}

\newcommand{\code}[1]{\texttt{#1}}

\newcommand{\secref}[1]{Sec.\,\ref{#1}}
\newcommand{\tabref}[1]{Tab.\,\ref{#1}}
\newcommand{\figref}[1]{Fig.\,\ref{#1}}

\newcommand{\ie}{i.e.,\xspace}
\newcommand{\eg}{e.g.,\xspace}
\newcommand{\cf}{cf.\xspace}

\newcommand{\wrt}{w.r.t.\xspace}

\usepackage{listings}
\begin{acronym}
	\acro{BN}{Bayesian Network}
	\acro{BLSTM}{Bidirectional Long Short Term Memory}
	\acro{DNN}{Deep Neural Network}
	\acro{DTMC}{Discrete-time Markov Chain}
	\acro{LSTM}{Long Short-Term Memory}
	\acro{GMM}{Gaussian Mixture Model}
	\acro{NN}{Neural Network}
	\acro{PST}{Probabilistic Suffix Tree}
	\acro{RF}{Random Forest}
	\acro{SVD}{Singular Value Decomposition}
	\acro{SVM}{Support Vector Machine}
	\acro{TA}{Timed Automata}

	\acro{CPS}{cyber-physical system}
	\acro{HMI}{Human-Machine Interface}
	\acro{PLC}{Programmable Logic Controller}
	\acro{RTU}{Remote Terminal Unit}
	\acro{ICS}{Industrial Control System}
	\acro{IDS}{Intrusion Detection System}
	\acro{IIDS}{Industrial Intrusion Detection System}

	\acro{DoS}{Denial of Service}
	\acro{DPI}{Deep Packet Inspection}
	\acro{IP}{Internet Protocol}
	\acro{MitM}{Machine-in-the-Middle}

	\acro{SWaT}{Secure Water Treatment}
	\acro{TEP}{Tennessee Eastman Process}
	\acro{WADI}{Water Distribution}

	\acro{PASAD}{Process-Aware Stealthy Attack Detector}
	\acro{TABOR}{Timed Automata and Bayesian netwORk}
	\acro{OOA}{Out Of Alphabet}
	\acro{SWIDE}{Sliding WIndow based on Differential sEgmentation}
	\acro{TPR}{True Positive Rate}
	\acro{FPR}{False Positive Rate}
	\acro{ODR}{Overall Detection Rate}

	\acro{IPAL}{Industrial Protocol Abstraction Layer}
\end{acronym}
\begin{document}

\title[IPAL: Breaking up Silos of Protocol-dependent and Domain-specific Industrial Intrusion Detection Systems]{IPAL: Breaking up Silos of Protocol-dependent and Domain-specific Industrial Intrusion Detection Systems}

\author{Konrad Wolsing}
\affiliation{%
   \institution{Fraunhofer FKIE}
   \country{}
   \city{}
}
\affiliation{%
   \institution{RWTH Aachen University}
   \country{}
   \city{}
}
\email{konrad.wolsing@fkie.fraunhofer.de}
\orcid{0000-0002-7571-0555}

\author{Eric Wagner}
\affiliation{%
   \institution{Fraunhofer FKIE}
   \country{}
   \city{}
}
\affiliation{%
   \institution{RWTH Aachen University}
   \country{}
   \city{}
}
\email{eric.wagner@fkie.fraunhofer.de}
\orcid{0000-0003-3211-1015}

\author{Antoine Saillard}
\affiliation{%
   \institution{RWTH Aachen University}
   \country{}
   \city{}
}
\affiliation{%
   \institution{Fraunhofer FKIE}
   \country{}
   \city{}
}
\email{antoine.saillard@rwth-aachen.de}
\orcid{0000-0002-8376-2726}

\author{Martin Henze}
\affiliation{%
   \institution{RWTH Aachen University}
   \country{}
   \city{}
}
\affiliation{%
   \institution{Fraunhofer FKIE}
   \country{}
   \city{}
}
\email{henze@cs.rwth-aachen.de}
\orcid{0000-0001-8717-2523}

\renewcommand{\shortauthors}{Wolsing et al.}

\begin{abstract}
The increasing interconnection of industrial networks %
exposes them to an ever-growing risk of cyber attacks.
To reveal such attacks early and prevent any damage, %
industrial intrusion detection searches for anomalies in otherwise predictable communication or process behavior.
However, current efforts %
mostly focus on specific domains and %
protocols, leading to a research landscape %
broken up into isolated silos.
Thus, existing approaches cannot be applied to other industries %
that would equally benefit from powerful detection. %
To better understand this issue, we survey 53 detection systems and %
find no fundamental reason for their narrow focus.
Although they are often coupled to specific industrial protocols in practice, many approaches could generalize to new industrial scenarios in theory.
To unlock this potential, %
we propose IPAL, our industrial protocol abstraction layer, to decouple intrusion detection from domain-specific industrial
protocols.
After proving IPAL's %
correctness in a reproducibility study of related work, we showcase its unique benefits %
by studying the generalizability of existing %
approaches to new datasets and conclude that they are indeed not restricted to specific domains or protocols and can perform outside their restricted silos.

\end{abstract}



\maketitle

\section{Introduction}
\label{sec:introduction}

Digitized industries play an essential role in today's society across various domains such as water treatment and distribution, power delivery, chemical processing, or manufacturing~\cite{humayed2017cyber}.
To realize their functionality, industries rely on (near) real-time communication of process values and commands using a large pool of specialized industrial protocols, such as Modbus or EtherNet/IP~\cite{2018_hiller_secure}. %
The increasing reliance of these protocols on the \acf{IP} to foster automation, remote control, and optimized processes, however, also moves traditionally air-gapped industrial networks closer to cybersecurity threats~\cite{humayed2017cyber}.
Indeed, a growing number of cyber incidents with detrimental environmental damage and risk to human life~\cite{ventures2019, osti_1505628} highlight the importance and urgency of adequately protecting industrial communication networks.

As a non-intrusive, retrofittable, and cheaply deployable security solution, \acp{IDS} offer a great additional layer of defense that is already well-established in traditional environments, such as office or data-center networks.
While these traditional \acp{IDS} (\eg Zeek~\cite{Zeek}) offer a certain benefit to industrial networks, they also reach their limitations:
Stealthy attacks %
differing just enough from normal operations to cause serious harm are hardly detectable by them~\cite{2016_urbina_limiting}.
Meanwhile, industrial protocol's rather deterministic and predictable nature offers unique %
opportunities for \emph{process-aware} or \emph{semantic} intrusion detection~\cite{giraldo2018survey}.
Hence, %
Industrial IDSs (IIDSs)\acused{IIDS} leveraging these regularities are necessary to protect industrial networks. %
Consequently, a large research community has gathered~\cite{2016_urbina_limiting,2020_olowononi_resilient,giraldo2018survey,ding2018survey,kaouk2019review,hu2018survey,loukas2019taxonomy,ramotsoela2018survey}, proposing specialized solutions for a wide variety of industrial scenarios. %

However, although different industrial domains exhibit similar predictable communication and process patterns, research on \acp{IIDS} is highly tailored to specific domains and communication protocols.
While technical and engineering reasons for such tight coupling might exist, we find this strong interdependence rather surprising and contrary to intuition.
From an \ac{IIDS} perspective, all industrial protocols essentially exhibit the same distinct characteristics:
They primarily exchange sensed data and commands %
using a small set of well-defined communication patterns~\cite{wolsing2020facilitating}.
Additionally, while specifics of benign and malicious behavior change between scenarios, these can mostly be trained prior to deployment. %

Although there exists no fundamental conceptual reason for the dependence of \acp{IIDS} on specific industrial protocols, \acp{IIDS} still specialize to few protocols, limiting their applicability to single scenarios.
Moreover, besides contrary claims~\cite{das2020anomaly,fovino2010modbus,inoue2017anomaly,kim2019anomaly}, it is seldom tested whether a specific \ac{IIDS} even works outside the precise scenario and industrial protocol it was developed for.
Consequently, the overall progress in research on \acp{IIDS} is slowed down due to niche solutions that do not improve on prior work from other industrial fields, wasting tremendous advancements in detecting cyber attacks on industrial systems.

\textbf{Contributions.} To break up the prevalent research silos of protocol-dependent and domain-specific \acp{IIDS} and ultimately transfer the advancements of \ac{IIDS} solutions to new industrial scenarios, we make the following contributions in this paper:

\begin{itemize}[noitemsep,topsep=5pt,leftmargin=9pt]
	\item %
	Beginning with a survey of 53 \acp{IIDS} from related work, we identify the characteristics of industrial communication they rely on to detect anomalies and attacks.
	Thus, we \emph{theoretically} identify the potential for protocol-independent and domain-agnostic intrusion detection across existing \acp{IIDS}~(\secref{sec:abstract-industrial-communication}).

	\acused{IPAL}
	\item To turn these results into \emph{practice}, we introduce the \acs{IPAL} industrial protocol abstraction layer.
	\ac{IPAL} sits between industrial protocols and \acp{IIDS}, captures all aspects relevant for intrusion detection in a unified representation, and thus facilitates protocol-independent and widely applicable \acp{IIDS}~(\secref{sec:ipal}).

	\item We showcase the practical \emph{applicability and correctness} of \ac{IPAL} in a reproducibility study of eight \acp{IIDS} from related work.
	We show that \ac{IPAL} provides all required information, and we simultaneously reproduce scientific results as an independent party, an important but often neglected step in research~(\secref{sec:reproducibility}).

	\item We illustrate the various \emph{benefits for \ac{IIDS} research} of \ac{IPAL} in three case studies.
	With \ac{IPAL}, we are able to compare different \acp{IIDS}, even from opposing research branches, study to what extent they generalize to new scenarios, and, for the first time, transfer the advancements of \acp{IIDS} %
	to new protocols and domains (\secref{sec:casestudies}).
\end{itemize}

\textbf{Availability Statement.}
To let \ac{IPAL} become the foundation of future work on industrial intrusion detection, we make our source code available.
This includes tools to transcribe industrial protocols~\cite{GitIPALTranscriber} and existing datasets into \ac{IPAL}~\cite{GitIPALDatasets}, as well as \acp{IIDS} implementations from our reproducibility study~\cite{GitIPALIIDS} (to the extent permitted by licenses or authors).

\section{Industrial Intrusion Detection}
\label{sec:iids}

As a foundation for our work, we begin by recapitulating the core idea behind intrusion detection and what makes it special in industrial networks.
Based on this, we identify key limitations of industrial intrusion detection research that prevent the broad application of improvements achieved for isolated industrial scenarios.

\subsection{Intrusion Detection}
\label{sec:iids:traditional}

Intrusion (or anomaly) detection is the art of automatically uncovering cyber attacks or other suspicious activity by passively monitoring a system's behavior, e.g., \wrt communication~\cite{stallings2015computer,wang2009ids}.
The core assumption behind \acfp{IDS}, which implement intrusion detection, is that cyber attacks lead to distinctively different (anomalous) system behavior than benign activity.
As their passive nature allows retrofitting them to existing deployments easily, \acp{IDS} are popular to complement preventive security measures, \ie integrity or authentication.

Traditionally, \acp{IDS} were designed for office, server, or data-center networks with largely fluctuating traffic and usage patterns~\cite{sommer2010outside}.
Even though this makes it challenging to define benign activity comprehensively, there is a decent understanding of how individual attacks are performed on these networks, as attacks and attack patterns are typically used across many different networks, e.g., by widely-used malware.
Thus, standard methods to reliably detect such (known) attacks are signature or rule-based approaches, which compare user and network behavior against a list of attack indicators, with well-established implementations such as YARA~\cite{YARA}, Suricata~\cite{Suricata}, Zeek~\cite{Zeek} (previously Bro~\cite{sommer2003bro}), and Snort~\cite{roesch1999snort}.

\subsection{Promises of Industrial Intrusion Detection}
\label{sec:iids:background}

Intrusion detection is particularly attractive for \emph{industrial} networks, as retrofitting other (active) security measures often results in costly hard- or software modifications, major downtime, or the need to redo acceptance tests. %
W.r.t attack and system behavior, however, %
we observe exactly opposite conditions for intrusion detection in industrial networks:
While rather deterministic industrial tasks lead to more predictable communication patterns~\cite{2018_hiller_secure}, attacks are often more subtle and scenario-specific, severely limiting predictability and thus signature or rule-based approaches.

Most importantly, attacks against industrial networks are often specifically designed for a single or few targets and hardly generalize across  domains (as, \eg malware does), making it hard to catch them by predefined rules.
Furthermore, industrial settings are susceptible to subtle attacks~\cite{2016_urbina_limiting,zemanek2022powerduck} where attackers send legitimate traffic (as allowed by rules) but at the ``wrong'' time to trip the system into a dangerous state.
Thus, \acp{IDS} designed for typical IT networks cannot simply be transferred to industrial settings~\cite{zhou2015multimodal}.

Still, industrial scenarios provide unique opportunities to detect intrusion and anomalies through different means, e.g., the tight coupling of industrial processes such as water treatment, power delivery, chemical processing, or manufacturing, with the physical environment via sensors and actuators.
To this end, \emph{process-aware} or \emph{semantic} \acp{IIDS} incorporate information from the physical and/or communication level to uncover highly specialized and subtle attacks and anomalies. %
Therefore, \ac{IIDS} approaches typically ``learn'' a model of benign behavior for a specific facility and alert deviations from this model.
Such models may incorporate additional system knowledge, e.g., that certain actuator combinations never occur, resulting in an \ac{IIDS} not requiring individual attack knowledge.

\subsection{Limitations of Current IIDS Research}
\label{sec:iids:landscape}

Given these promises of \acp{IIDS} to detect sophisticated and safety-critical attacks on industrial networks and processes, a plethora of different research fields for \acp{IIDS} in various industrial domains has been established~\cite{giraldo2018survey,2016_urbina_limiting, 2020_olowononi_resilient}, including industrial control systems~\cite{ding2018survey,kaouk2019review,hu2018survey}, intra-vehicular communication~\cite{loukas2019taxonomy}, or water treatment~\cite{ramotsoela2018survey}.
Across these various research fields, we find that individual \acp{IIDS} are tailored to specific domains and communication protocols.
Consequently, advancements made for one scenario often cannot generalize to other industrial protocols or domains, limiting their widespread use, slowing down innovation, and the overall progress in securing industrial networks.
We identify two core problems that lead to these limitations of current \ac{IIDS} research.

First, compared to traditional computer networks, we observe a huge \emph{heterogeneity} in industrial domains, each with unique and custom communication protocols for legacy reasons.
Thus, current \acp{IIDS} generally focus on one specific domain and industrial protocol combination, presenting yet another (novel) approach in isolation without building on top of prior research findings.
E.g., two similar approaches independently propose to use Probabilistic Suffix Trees for Modbus~\cite{yoon2014communication} and IEC-104~\cite{lin2018understanding}, highlighting that work is done twice across different industrial scenarios.

Second, \ac{IIDS} research across different industrial domains suffers from an \emph{evaluation bias}, where vast inconsistencies in used evaluation methodologies make comparisons across publications hardly possible~\cite{giraldo2018survey,2016_urbina_limiting,kus2022false}.
Moreover, the use of only a few datasets (\eg SWaT~\cite{goh2016dataset}) or many private ones steers the collective research efforts in a direction where solutions are optimized and biased towards specific scenarios.
This problem is already known for traditional \acp{IDS}~\cite{panigrahi2018detailed} but is even more severe in industrial settings, where systems expose very narrow yet domain-specific behavior~\cite{2020_turrin_datasets}.

Consequently, although different approaches claim broader applicability (without proof)~\cite{das2020anomaly,fovino2010modbus,inoue2017anomaly,kim2019anomaly}, the actual generalizability of proposed \acp{IIDS} to new settings within the same domain or across domains remains mostly unexplored~\cite{erba2020no,kus2022false}.
Thus, the various research streams in the area of \acp{IIDS} are indeed disjoint.
As a result, the overall research field is slowed down, as new approaches are designed for niches without improving on previous research and are biased to datasets from specific scenarios.
It thus remains unknown to what extent the tremendous advancements in the detection of cyber attacks on industrial systems can or cannot be generalized and transferred from their isolated niches to a broader scale and thus sustainably increase security for many industrial networks.

\section{The Case for Protocol-Independent Industrial Intrusion Detection}
\label{sec:abstract-industrial-communication}

\begin{figure}[t]
	\centering
	\includegraphics[width=\columnwidth]{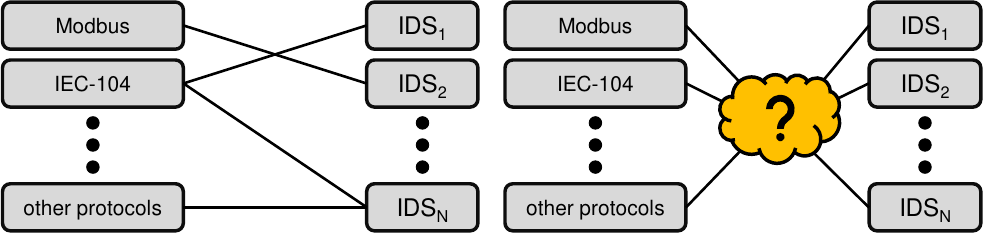}
	\begin{subfigure}{0.48\columnwidth}
		\caption{The current state-of-the-art requires a tailored implementation of \acp{IIDS} for each protocol.}
		\label{fig:idea_old}
	\end{subfigure}%
	\hfill
	\begin{subfigure}{0.48\columnwidth}
		\caption{Protocol-independent \acp{IIDS} would allow separating protocol parsing and IDS development.}
		\label{fig:idea_new}
	\end{subfigure}
	\vspace{-.5em}
	\caption{To overcome the limitations of current \ac{IIDS} research, protocol-independent \acp{IIDS} promise to break up the strong interdependence between \ac{IIDS} approaches and the specific industrial protocol and domain they operate in.}
	\label{fig:idea}
\end{figure}

Currently, the \ac{IIDS} research landscape is scattered across heterogeneous industrial domains (cf.~\secref{sec:iids:landscape}), leading to a situation as depicted in~\figref{fig:idea_old}, where \acp{IIDS} are developed in isolation for specific industrial protocols and domains.
Consequently, research on \acp{IIDS} is limited \wrt generalizability, evaluation scenarios, widespread use, and coherence between research streams.

To overcome these limitations, in this paper, we make a case for \emph{protocol-independent \acp{IIDS}}, as shown in~\figref{fig:idea_new}, where \acp{IIDS} do not have to care about the underlying protocol anymore while simultaneously being transferable to various scenarios with little to no additional effort.
Our proposal for protocol-independence in \acp{IIDS} is motivated by two observations:
First, for the purpose of intrusion detection, essentially all industrial protocols show similar functionality, i.e., the exchange of sensor readings and actuator commands.
Second, in principle, the underlying detection approaches of \acp{IIDS} work similarly as they each know or learn a model of a specific industrial process's behavior and detect anomalies by comparing their models to the monitored physical process.

Given these observations, we postulate that a large portion of existing work is not necessarily bound to the specifics of the industrial domain and protocol they are designed for and may generalize well to other domains and protocols.
Thus, we are convinced that tackling the diversity in industrial protocols is key to breaking up the isolated silos in the \acp{IIDS} research landscape and hence allows the wide-ranging application of \acp{IIDS} across domains and protocols.

To realize this goal, we first highlight the benefits of protocol-independent \acp{IIDS} for the research community (\secref{sec:abstract-industrial-communication:benefits}) before we survey 53 state-of-the-art \ac{IIDS} approaches to analyze their potential for generalizing across industrial domains and protocols (\secref{sec:iids-survey}).

\subsection{Benefits of Protocol-Independent IIDSs}
\label{sec:abstract-industrial-communication:benefits}

Although \acp{IIDS} are specifically tailored to a single industrial protocol and domain nowadays, this does not imply that their benefits are necessarily restricted to a single combination.
\emph{Protocol-independent \acp{IIDS}} (cf.~\figref{fig:idea_new}) instead promise to operate detached from specific industrial protocols or domains and thus exhibit several advantages.

First, this would allow applying \acp{IIDS} to multiple industrial protocols within the same domain.
E.g., power grids that historically rely on different protocols such as IEC-104 (covered by \acp{IIDS} \cite{lin2019timing,lin2017timing,ferling2018intrusion,lin2018understanding}) or DNP3 (covered by \cite{fovino2010modbus,radoglou2020diderot}) would no longer require separate \acp{IIDS}. %
Moreover, if the underlying industrial protocol could be exchanged, the potential to generalize detection methodologies across domains not initially thought of would become directly accessible.

Also, by designing \acp{IIDS} with protocol-independence in mind, developers no longer have to extract relevant data from a specific industrial protocol in a time-consuming and error-prone process.
Instead, this process can be left to domain experts who know the ins and outs of specific network protocols, which has already proved beneficial for prominent intrusion detection datasets, e.g. SWaT~\cite{goh2016dataset} providing pre-processed process data from network traffic.

Finally, it becomes easier to reproduce results, validate findings in different scenarios, and fairly compare \acp{IIDS} based on common metrics, thus constituting to sound scientific experiments~\cite{giraldo2018survey,uetz2021socbed}.
Protocol-independence allows to efficiently validate results from prior work and expand upon them, a step that is often neglected in research~\cite{bajpai2019reproducibility,uetz2021socbed}, especially when the original scenario is unavailable, flawed, or too narrow in scope.

Consequently, decoupling \acp{IIDS} from the underlying industrial protocol promises to address the transferability of \acp{IIDS} and their generalizability to new scenarios. It can also streamline future research efforts on \acp{IIDS} by enabling extensive evaluations, validations, or removing potential biases with regard to specific protocols or domains.
However, so far, it remains open how big the potential for realizing protocol-independent \acp{IIDS} actually is.

\subsection{Potential for Protocol-Independent IIDSs}
\label{sec:iids-survey}

To identify the potential for protocol-independent \acp{IIDS}, we first study to which extent existing \acp{IIDS} can generalize to new industrial protocols and domains, which is directly related to the protocol and process information they operate on. %
However, while existing surveys identify the need for generalizing \acp{IIDS}~\cite{erba2020no}, %
they do not investigate whether existing work shows potential in this regard.
Bridging this gap, we survey 53 \acp{IIDS} %
covering a wide range of detection methods, industrial protocols, and domains.
We specifically focus on identifying which information they use for detection. %

As our goal is to identify the potential for protocol-independent \acp{IIDS}, we specifically target \emph{process-aware} or \emph{semantic} \acp{IIDS} (leveraging knowledge about the physical process), as these are mostly independent of the used protocol.
Contrary, we exclude \acp{IIDS} leveraging specific characteristics of physical communication~\cite{kneib2018scission,kneib2020easi}.

\textbf{Survey Structure.}
In the following, we structure and discuss our survey, as depicted in \tabref{tab:iids-summary}, along two broad categories of process-aware and semantic intrusion detection:
\emph{Communication-based} approaches (\secref{sec:iids-survey-network}) which incorporate characteristics of industrial network traffic and \emph{process state-aware} approaches (\secref{sec:iids-survey-state}) restricted to the current (physical) state of an industrial setting.

For each approach, we identify the input format, \ie network packets (P) or %
aggregated process states (S).
We complement this with %
information an \ac{IIDS} operates on, \ie \textit{(i)} timing information of exchanged packets, \textit{(ii)} knowledge about communication partners and type of communication, as well as \textit{(iii)} which values are communicated, also including the process state.
We identify whether additional packet features are required (\eg packet lengths), whether an externally provided process model is required, as well as whether it trains on anomalous (A) or benign-only (B) traffic.
Finally, we list the number of evaluation scenarios, the number of \acp{IIDS} a novel approach compares to, the intended industrial domain and protocol, as well as the availability of source code and evaluation dataset.

\begin{table*}[!htbp]
	\centering \footnotesize

	\newcolumntype{C}[1]{>{\centering}m{#1}}
	\newcommand{\rowend}[1]{\rowcolor{#1} \cellcolor{white}}
	\newcommand{\litref}[1]{} %
	\newcommand*\circled[1]{\tikz[baseline=(char.base)]{\node[shape=circle,draw,inner sep=0pt] (char) {#1};}}
	
	\newcommand{\water}{\includegraphics[height=2.1mm]{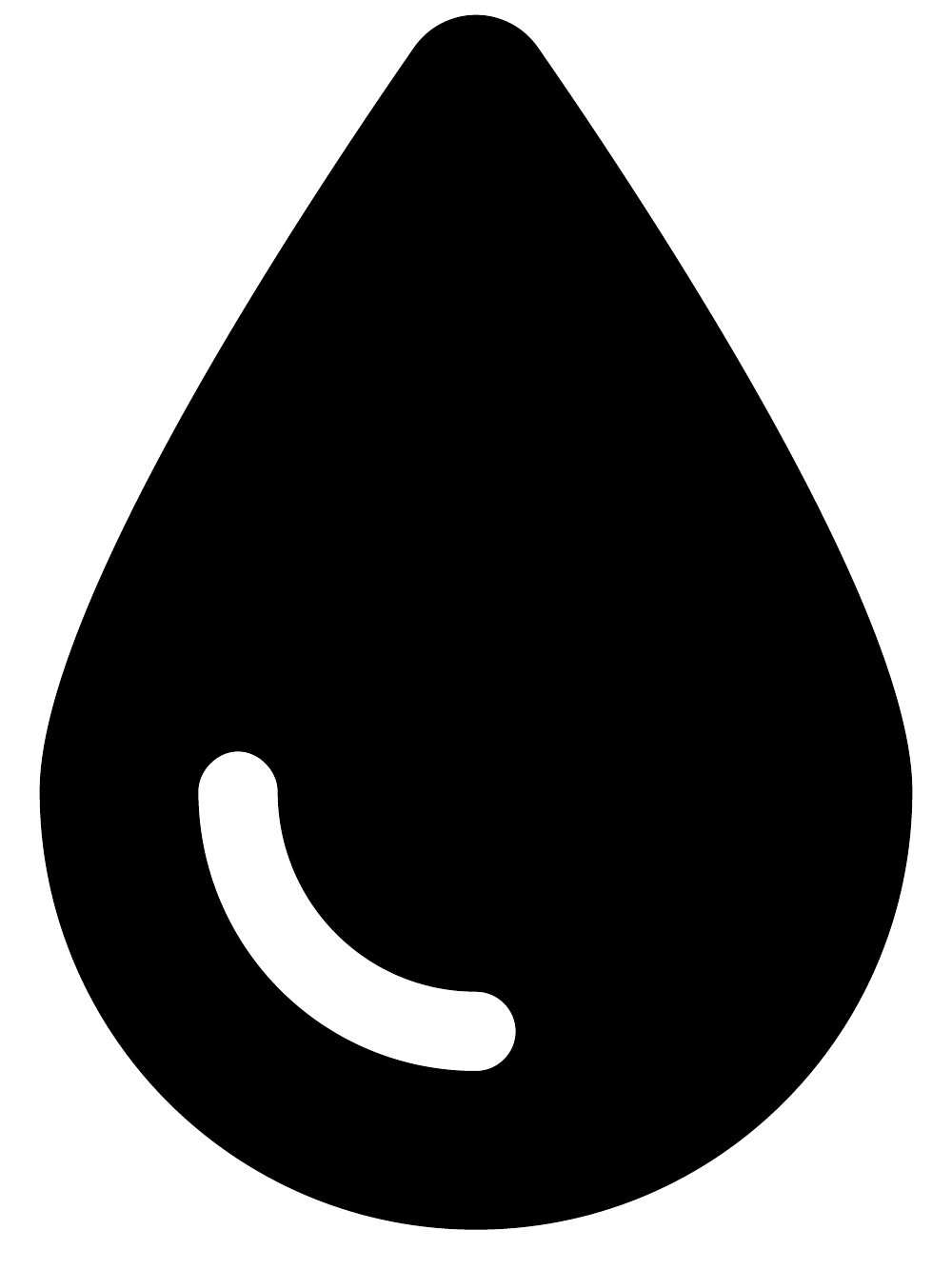}} %
	\newcommand{\medical}{\includegraphics[height=2.1mm]{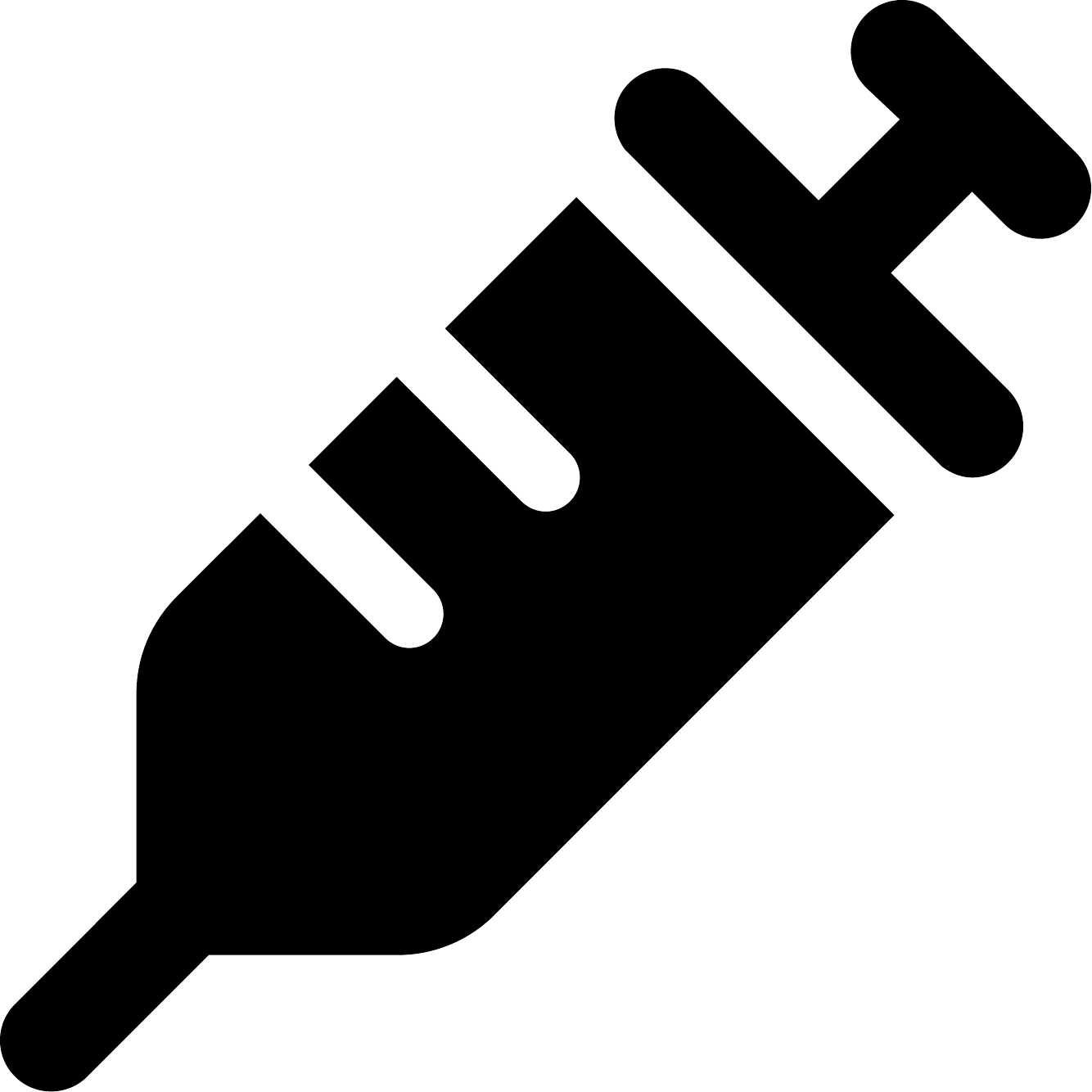}} %
	\newcommand{\dist}{\includegraphics[height=2.1mm]{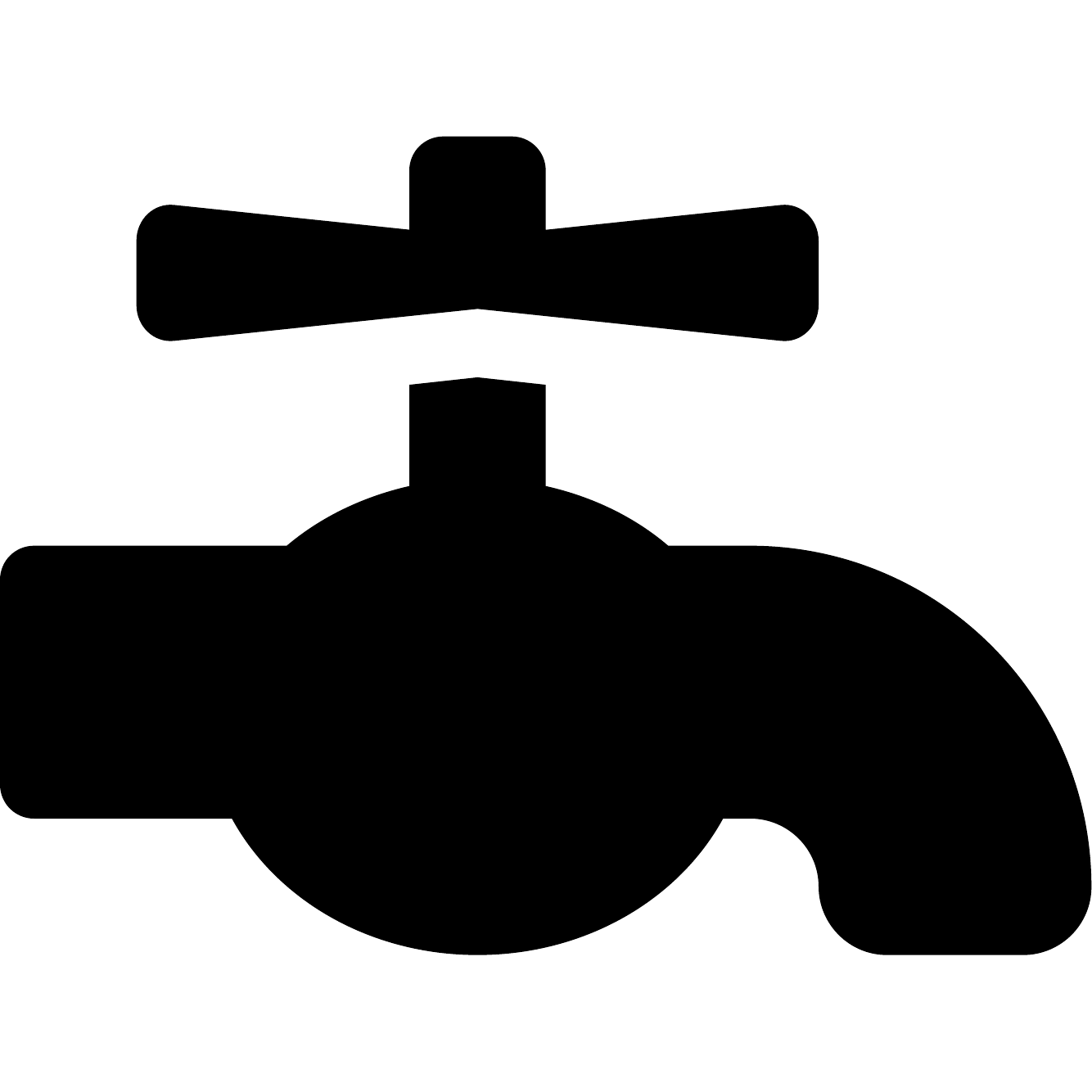}} %
	\newcommand{\power}{\includegraphics[height=2.1mm]{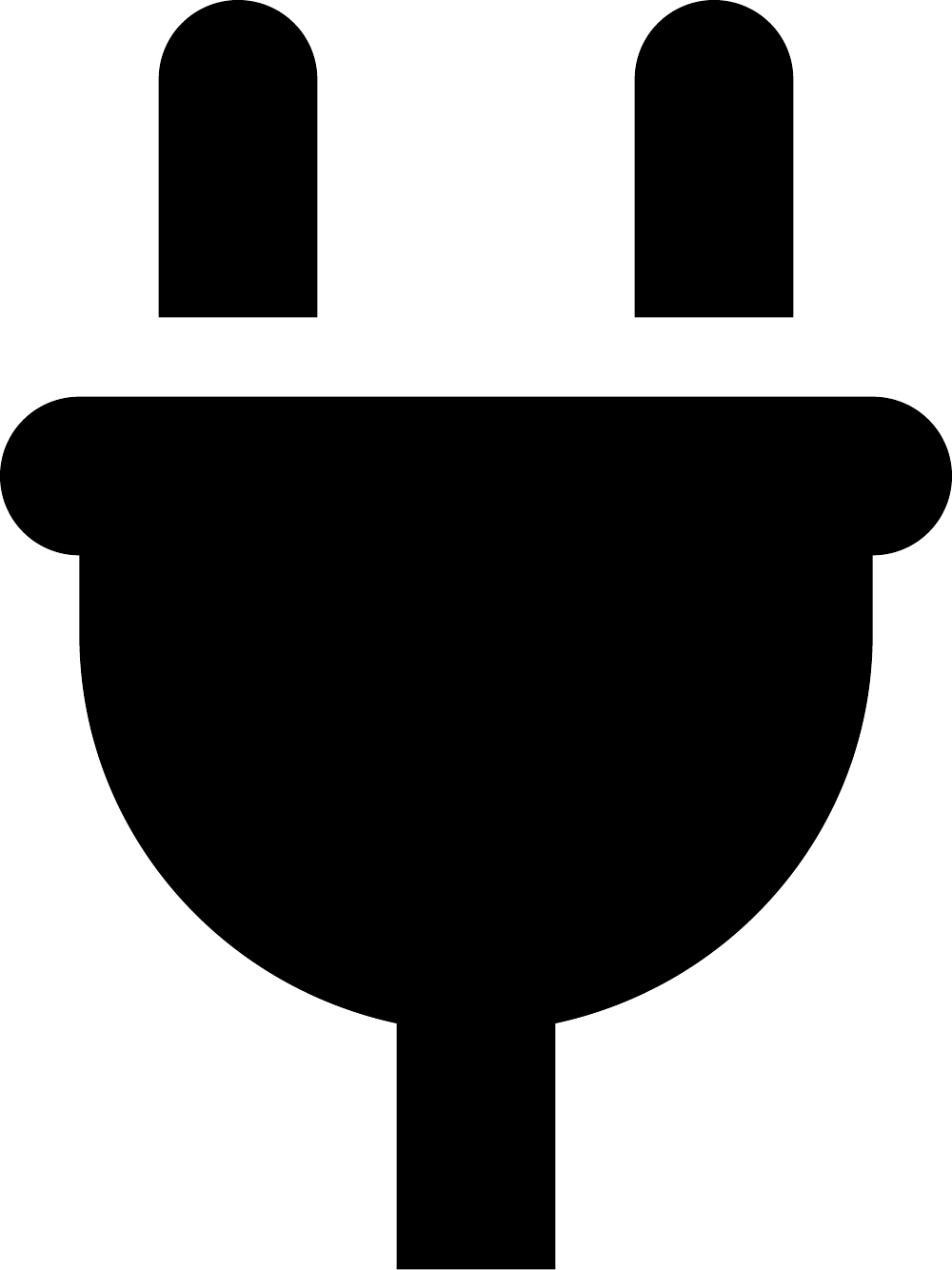}} %
	\newcommand{\gas}{\includegraphics[height=2.1mm]{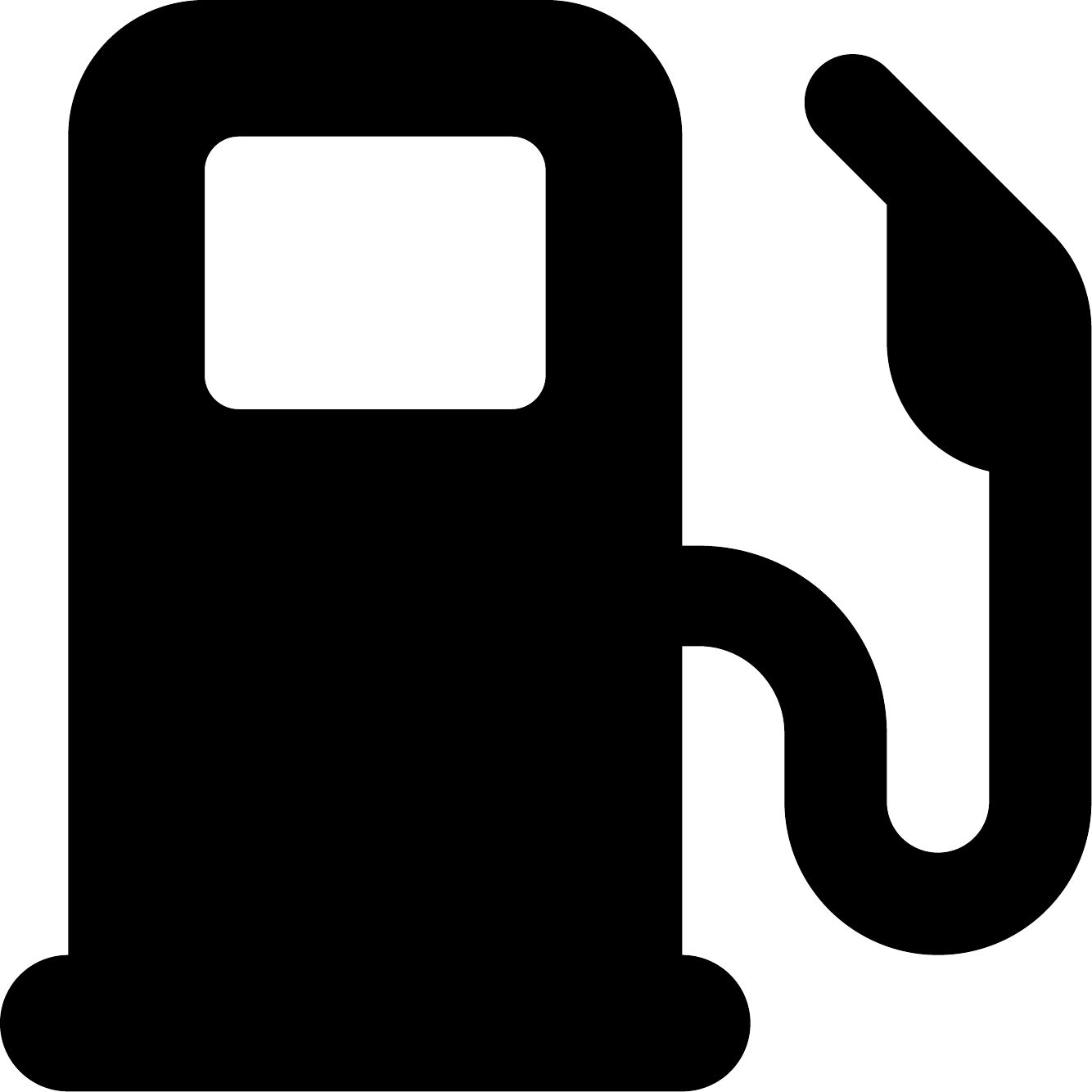}} %
	\newcommand{\chemical}{\includegraphics[height=2.1mm]{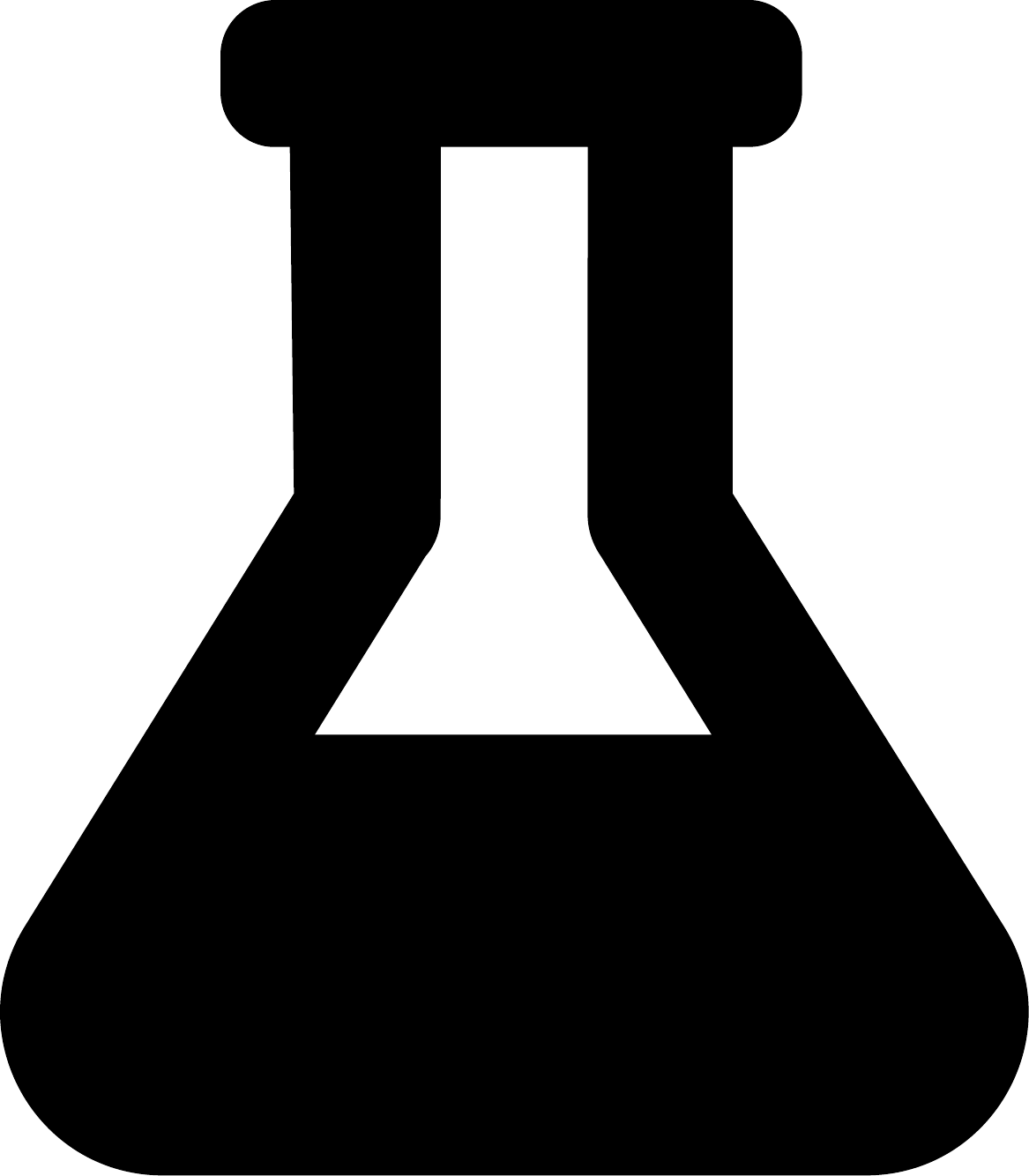}} %
	\newcommand{\manufacturing}{\includegraphics[height=2.1mm]{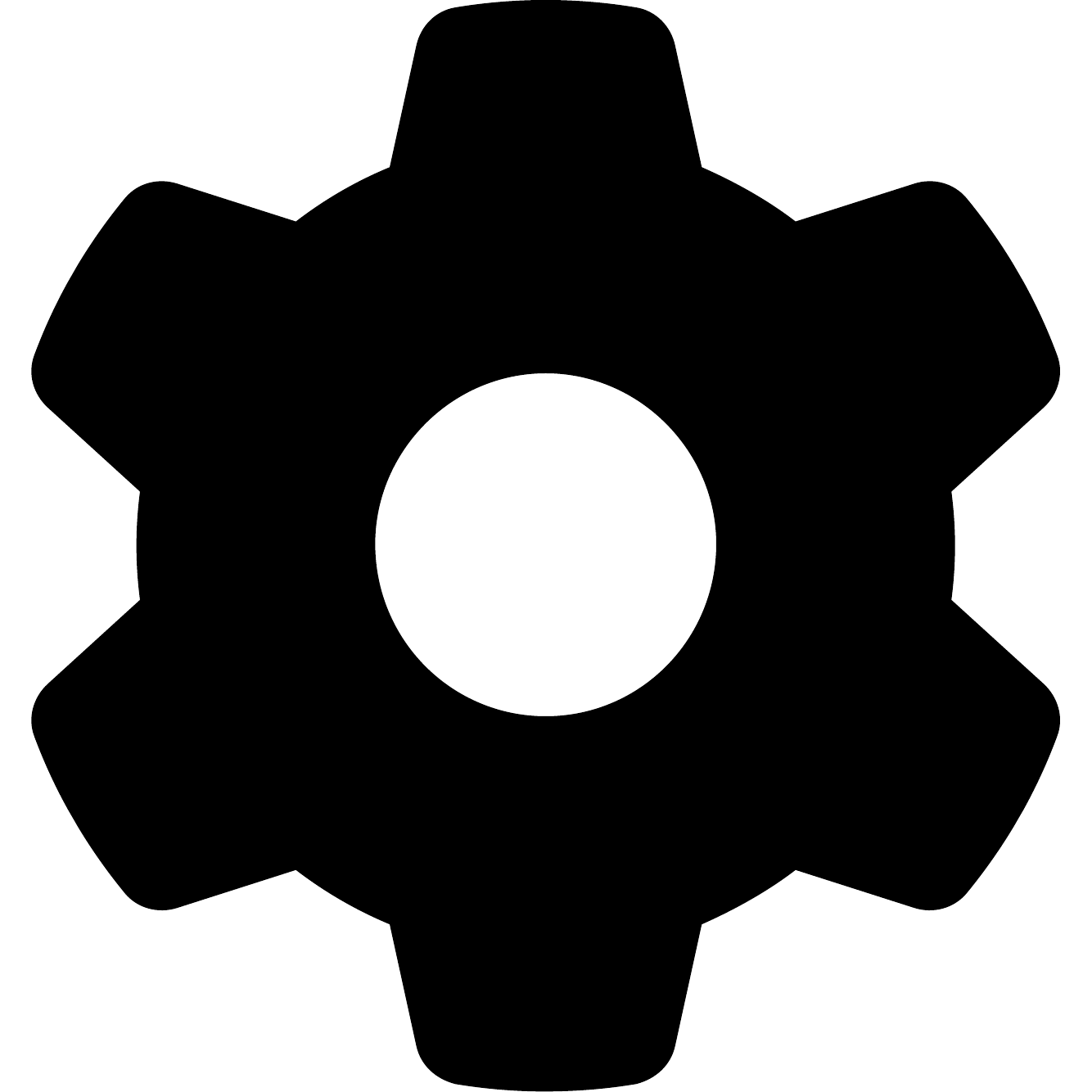}} %
	\newcommand{\transport}{\includegraphics[height=2.1mm]{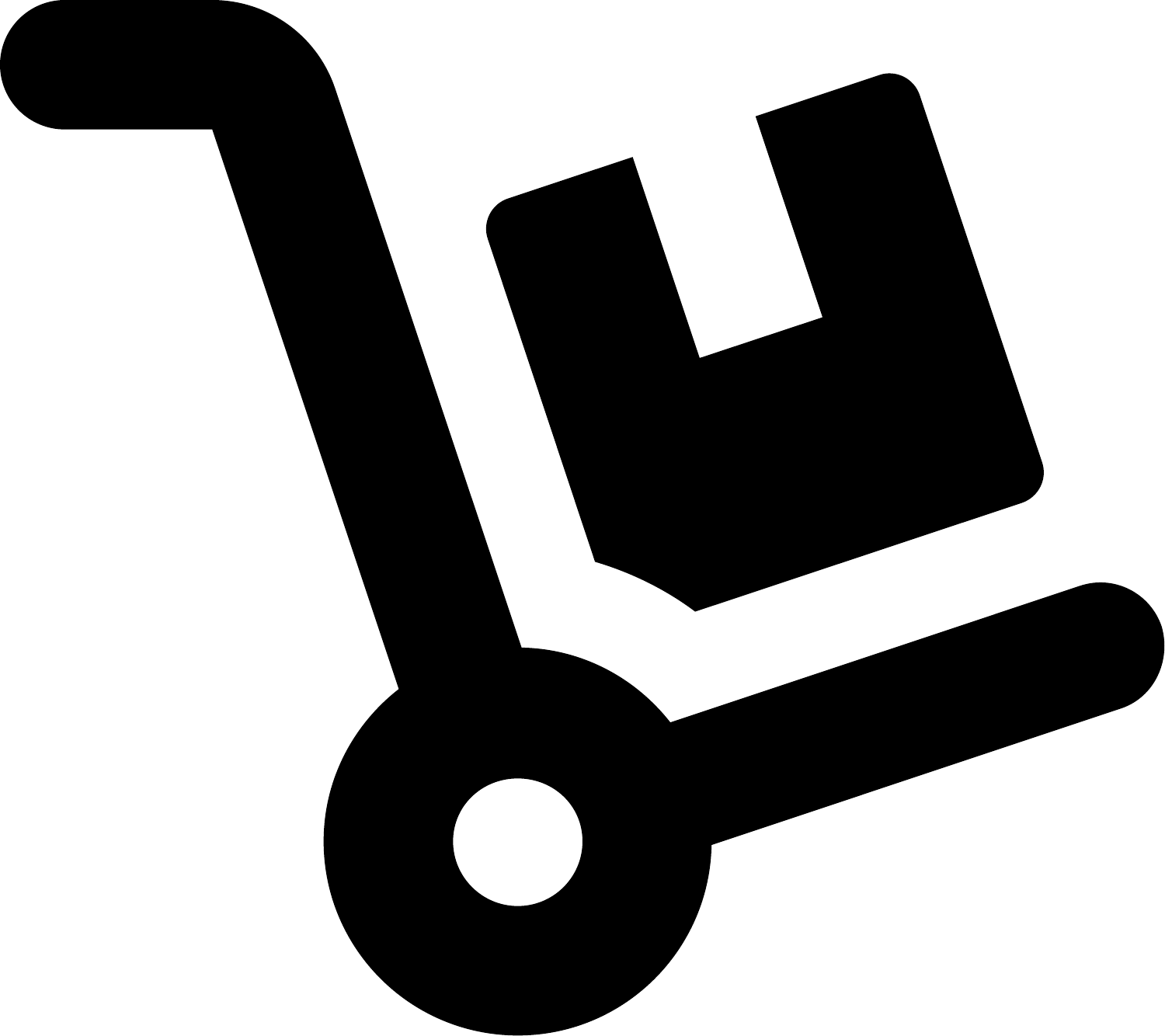}} %
	\newcommand{\synthetic}{\includegraphics[height=2.1mm]{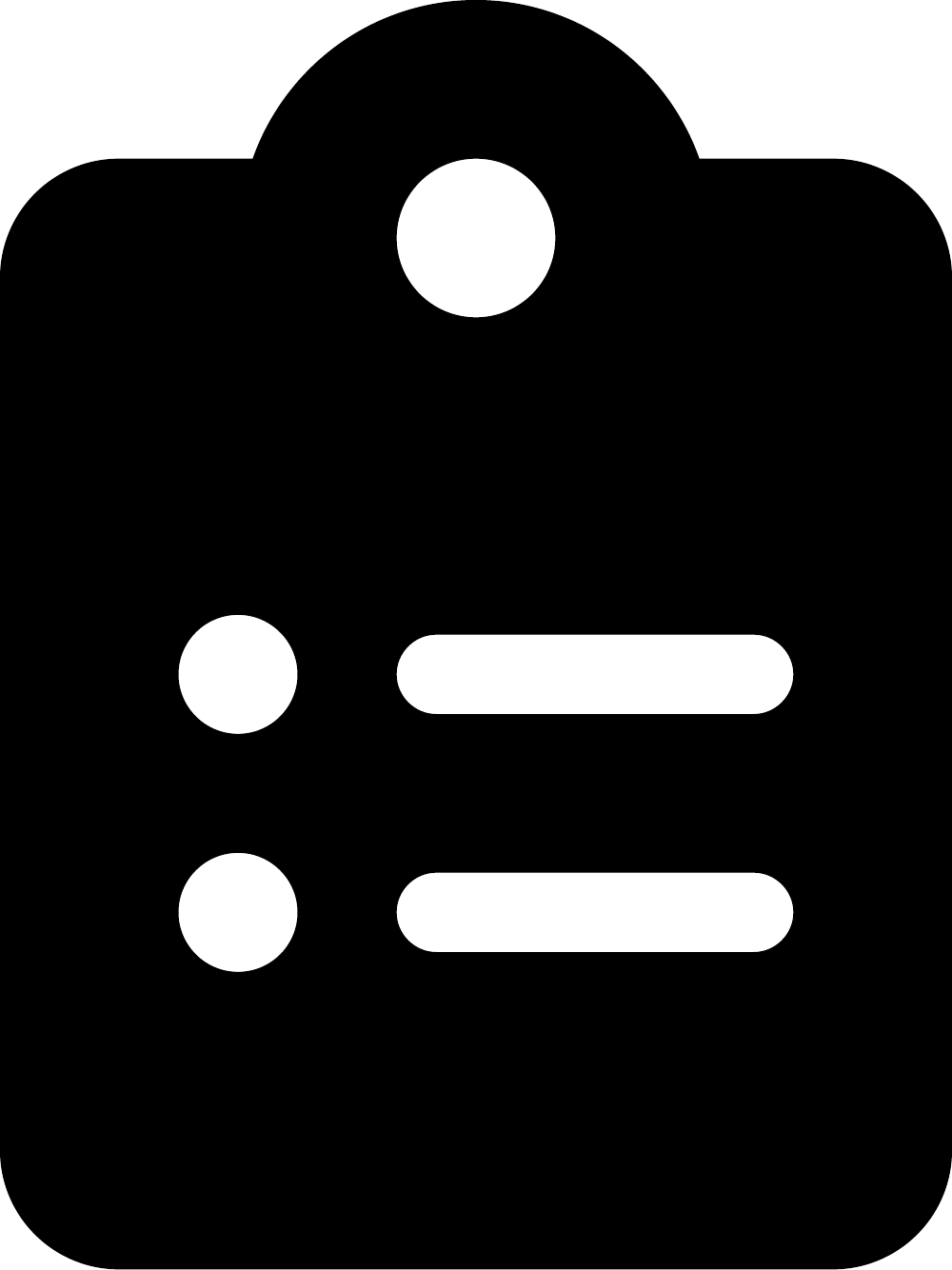}} %

	\makeatletter
	\def\adl@drawiv#1#2#3{%
		\hskip.5\tabcolsep
		\xleaders#3{#2.5\@tempdimb #1{1}#2.5\@tempdimb}%
		#2\z@ plus1fil minus1fil\relax
		\hskip.5\tabcolsep}
	\newcommand{\cdashlinelr}[1]{%
		\noalign{\vskip\aboverulesep
			\global\let\@dashdrawstore\adl@draw
			\global\let\adl@draw\adl@drawiv}
		\cdashline{#1}
		\noalign{\global\let\adl@draw\@dashdrawstore
			\vskip\belowrulesep}}
	\makeatother

	\begin{tabularx}{.95\textwidth}{ m{.1cm}>{\raggedright\arraybackslash} m{.1cm}>{\raggedright\arraybackslash} C{2.65cm} m{.4cm} >{\raggedright\arraybackslash}C{2.4cm} m{.06cm}m{.06cm}m{.06cm}m{.06cm}m{.06cm}m{.06cm}m{.06cm}m{.06cm}m{.25cm}
		m{.06cm}m{.17cm}m{.06cm}m{.06cm}m{.6cm}m{2.35cm}}
		
		& 
		& \vspace{13mm} \textbf{Publication(s)}
		& \vspace{13mm} \textbf{Year}
		& \vspace{13mm} \textbf{Detection Methodology} &
		\rotatebox[origin=t,]{50}{\parbox{2cm}{\textbf{Representation}}} &
		\rotatebox[origin=t,]{50}{\parbox{2cm}{\textbf{Timing}}} &
		\rotatebox[origin=t,]{50}{\parbox{2cm}{\textbf{Comm. Entities}}} &
		\rotatebox[origin=t,]{50}{\parbox{2cm}{\textbf{Message Type}}} &
		\rotatebox[origin=t,]{50}{\parbox{2cm}{\textbf{Comm. Values}}} &
		\rotatebox[origin=t,]{50}{\parbox{2cm}{\textbf{Process State}}} &
		\rotatebox[origin=t,]{50}{\parbox{2cm}{\textbf{Packet Features}}} &
		\rotatebox[origin=t,]{50}{\parbox{2cm}{\textbf{External Model}}} &
		\rotatebox[origin=t,]{50}{\parbox{2cm}{\textbf{Trains on}}} &

		\rotatebox[origin=t,]{50}{\parbox{2cm}{\textbf{Open Source}}} &
		\rotatebox[origin=t,]{50}{\parbox{2cm}{\textbf{Public Dataset}}} &
		\rotatebox[origin=t,]{50}{\parbox{2cm}{\textbf{Eval. Scenarios}}} &
		\rotatebox[origin=t,]{50}{\parbox{2cm}{\textbf{Compares to}}} &
		\rotatebox[origin=t,]{50}{\parbox{2cm}{\textbf{Domain(s)}}} &
		\vspace{13mm} \textbf{Protocol(s)} \\
		 \toprule %

		&& Valdes \etal \cite{valdes2009communication} \litref{63} & 2009 &  Flow Periodicity & P & \good & \good & \bad & \bad & \bad &\good & \bad & B &\bad & \bad & 1 & 0 & \synthetic & Modbus \\
		&& Barbosa \etal \cite{barbosa2012towards} \litref{65} &2012 &  Periodicity & P & \good & \good & \bad & \bad & \bad & \bad & \bad & B & \bad & \bad & 1 & 0 & \water & -- \\
		&& Ponomarev \etal \cite{ponomarev2015industrial} \litref{25} &2016 & Telemetry& P & \good & \good & \bad & \bad & \bad & \good & \bad & A/B & \bad & \bad & 1 & 0 & \synthetic & Modbus \\
		&& Lin \etal \cite{lin2017timing} \litref{23} & 2017& Inter arrival time & P & \good & \good & \good & \good & \bad & \bad & \bad & B & \bad & \ok & 3 & 0 & \power & Modbus,\,IEC-104,\,S7 \\
		&\multirow[c]{-7.32}{*}{\rotatebox[origin=t, ]{90}{\parbox{2.6cm}{\centering\textbf{Timings}}}}& Lin \etal \cite{lin2019timing} \litref{48} &2019 &  Inter-arrival time & P & \good & \good & \good & \bad & \bad & \bad & \bad & B & \bad & \ok & 1 & 0 & \power & IEC-104\\ %
		\cdashlinelr{2-20}
		
		&& Goldenberg \etal \cite{goldenberg2013accurate} \litref{22} & 2013 & DFA & P & \good & \good & \good & \ok & \bad & \bad & \bad & B & \bad & \bad & 1 & 0 & \manufacturing & Modbus \\
		&&Yoon \etal \cite{yoon2014communication} \litref{42} & 2014 & PST & P & \ok & \good & \good & \good & \bad & \bad & \bad & B & \bad & \bad & 2 & 0 & \synthetic & Modbus \\
		&& Caselli \etal \cite{caselli2015sequence,caselli2015modeling} \litref{17} &2015 & DTMC & P & \good & \good & \good & \good & \bad & \bad & \bad & B & \bad & \bad & 1 & 0 & \water & Modbus \\
		&& Ferling \etal \cite{ferling2018intrusion} \litref{21} & 2018& DTMC & P & \good & \good & \good & \good & \bad & \bad & \bad & B & \good & \bad & 1 & 0 & \gas & IEC-104 \\ %
		&&Lin \etal \cite{lin2018understanding} \litref{28} & 2018 & PST & P & \good & \good & \good & \good & \bad & \bad & \bad & B & \ok & \bad & 2 & 0 & \power & IEC-104 \\
		&\multirow[c]{-6.1}{*}{\rotatebox[origin=t,]{90}{\centering\textbf{Sequences}}}& Yun \etal \cite{yun2018statistical} & 2018 & Nearest Neighbor & P & \good & \good & \bad & \bad & \bad & \bad & \bad & B & \bad & \bad & 1 & 0 & -- & -- \\ %
		\cdashlinelr{2-20}

		&& Shang \etal \cite{shang2016modbus} \litref{30} & 2016 & SVM & P & \good & \good & \good & \good & \bad & \good & \bad & A/B & \bad & \bad & 1 & 0 & \chemical & Modbus \\ %
		&& Feng \etal \cite{feng2017multi} \litref{20} & 2017 & LSTM & P & \good & \good & \good & \bad & \good & \good & \bad & B & \bad & \good$^M$ & 1 & 1 & \gas & Modbus \\
		&& Perez \etal \cite{perez2018machine} \litref{54} & 2018 & SVM, RF, BLSTM & P & \good & \good & \good & \bad & \good & \good & \bad & A/B & \good & \good$^M$ & 1 & 0 & \gas & Modbus \\
		&& Anton \etal \cite{anton2019anomaly} \litref{32} & 2019 & RF, SVM & P & \good & \good & \good & \bad & \good & \good & \bad & A/B & \bad & \good$^M$ & 1 & 0 & \gas & Modbus \\
		&& Chu \etal \cite{chu2019industrial} \litref{15} & 2019 & NN & P & \bad & \bad & \good & \bad & \bad & \good & \bad & A/B & \bad & \good$^M$ & 1 & 3 & \gas & Modbus \\
		&& Maglaras \etal \cite{maglaras2014intrusion} \litref{31} & 2019 & OCVM & P & \good & \ok & \bad & \bad & \bad & \good & \bad & B & \bad & \bad & 1 & 0 & -- & Modbus \\
		\multirow[c]{-20.5}{*}{\rotatebox[origin=t, ]{90}{\parbox{5.5cm}{\centering\textbf{Communication-based Approaches  (\secref{sec:iids-survey-network})}}}}&	\multirow[c]{-7.9}{*}{\rotatebox[origin=t,]{90}{\centering\textbf{Classifiers}}}& DIDEROT~\cite{radoglou2020diderot} \litref{71} & 2020 & Decision Tree, DNN & P & \good & \good & \bad & \bad & \bad & \good & \bad & A/B & \bad & \bad & 1 & 0 & \power & DNP3 \\ \toprule

		&& Carcano \etal \cite{carcano2011multidimensional} \litref{10} & 2011& Language & S & \bad & \bad & \bad & \bad & \good & \bad & \good & A & \bad & \bad & 1 & 0 & \power & Modbus \\
		&& Almalawi \etal \cite{almalawi2014unsupervised} \litref{2} & 2014& Outlier detection & S & \bad & \bad & \bad & \bad & \good & \bad & \bad & A/B & \bad & \ok & 3 & 0 & \water & Modbus \\
		&& Kiss \etal \cite{kiss2015clustering} \litref{39} &2015 & GMM & S & \bad & \bad & \bad & \bad & \good & \bad & \bad & B & \bad & \good$^T$ & 1 & 0 & \chemical & -- \\
		&& Pan \etal \cite{pan2015developing} & 2015 & Common Paths & S & \good & \bad & \bad & \bad & \good & \bad & \bad & A/B & \bad & \bad & 1 & 0 & \power & Modbus,\,IEEE C37.118\\ %
		&& SysDetect \cite{khalili2015sysdetect} \litref{38} &2015 & Frequent Itemsets & S & \bad & \bad & \bad & \bad & \good & \bad & \good & A/B & \bad & \bad & 1 & 0 & \manufacturing & -- \\
		&& Kong \etal \cite{kong2016temporal,jones2014anomaly,kong2014temporal} \litref{12,29,62} &2016 & Temporal logic & S & \good & \bad & \bad & \bad & \good & \bad & \bad & B & \ok & \bad & 2 & 0 & \transport & -- \\
		&& Adepu \etal \cite{adepu2016distributed} \litref{58} & 2016 & Invariants & S & \good & \bad & \bad & \bad & \good & \bad & \good & - & \bad & \good$^S$ & 1 & 0 & \water &  EtherNet/IP \\
		&& Feng \etal \cite{feng2019systematic} \litref{3} & 2019& Invariants & S & \bad & \bad & \bad & \bad & \good & \bad & \bad & B & \ok & \good$^S_W$ & 2 & 2 & \water \vspace{.001mm} \dist & EtherNet/IP \\
		&& Monzer \etal \cite{monzer2019model} \litref{11} & 2019& Rules & S & \good & \bad & \bad & \bad & \good & \bad & \good & B & \bad & \bad & 1 & 0 & \water & Modbus \\
		&\multirow[c]{-10.7}{*}{\rotatebox[origin=t,]{90}{\centering\textbf{Critical States}}}& Das \etal \cite{das2020anomaly} \litref{49} &2020 & Data Analysis & S & \ok & \bad & \bad & \bad & \good & \bad & \bad & A/B & \bad & \good$^S$ & 1 & 0 & \water & EtherNet/IP \\
		\cdashlinelr{2-20}

		&& Had\v{z}iosmanovi\'{c} \etal \cite{hadvziosmanovic2014through} \litref{16} &2014 & Autoregression & S & \good & \bad & \bad & \bad & \good & \bad & \bad & B & \ok & \bad & 2 & 0 & \water & Modbus \\
		&& Caselli \etal \cite{caselli2015sequence} \litref{17} & 2015& DTMC & S & \good & \bad & \bad & \bad & \good & \bad & \bad & B & \bad & \bad & 1 & 0 & \water & EtherNet/IP \\
		&& Ahmed \etal \cite{ahmed2017model} \litref{57}& 2017 & Kalman Filter & S & \good & \bad & \bad & \bad & \good & \bad & \bad & B & \bad & \bad & 1 & 0 & \dist & -- \\
		&& PASAD \cite{aoudi2018truth,aoudi2020scalable} \litref{13,41} &2018 & PCA & S & \good & \bad & \bad & \bad & \good & \bad & \bad & B & \good & \ok$^S_T$ & 3 & 1 & \water \hspace*{.1mm} \chemical & Modbus EtherNet/IP \\
		&& Choi \etal \cite{choi2018detecting} \litref{53} & 2018 & Control Invariants & S & \good & \bad & \bad & \bad & \good & \bad & \bad & B & \good & \bad & 1& 0 & \transport & -- \\
		&& Myers \etal \cite{myers2018anomaly} & 2018 & Petri-nets & S & \good & \bad & \bad & \bad & \good & \bad & \bad & B & \bad & \good & 1 & 0 & \manufacturing & S7 \\ %
		&& Kravchik \etal \cite{kravchik2018detecting} & 2018 & Neural Networks & S & \good & \bad & \bad & \bad & \good & \bad & \bad & B & \bad & \good & 1 & 3 & \water & EtherNet/IP \\ %
		&& TABOR \cite{lin2018tabor} \litref{8} & 2018 & TA, BN & S & \good & \bad & \bad & \bad & \good & \bad & \bad & B & \bad & \good$^S$ & 1 & 2 & \water & EtherNet/IP \\
		&& Anton \etal \cite{duque2019security} \litref{6} & 2019 & Matrix Profiles & S & \good & \bad & \bad & \bad & \good & \bad & \bad & B & \bad & \good$^S$ & 1 & 1 & \water & EtherNet/IP \\
		&& HybTester \cite{castellanos2019modular} \litref{7} &2019 & Hybrid-Automata & S & \good & \bad & \bad & \bad & \good & \bad & \bad & B & \bad & \good$^S$ & 1 & 0 & \water & EtherNet/IP \\
		&& Kim \etal \cite{kim2019anomaly} \litref{50} & 2019& Neural Networks & S & \good & \bad & \bad & \bad & \good & \bad & \bad & B & \good & \good$^S$ & 1 & 1 & \water & EtherNet/IP \\
		&& Denque Anton \cite{anton2020intrusion} \litref{45} & 2020& Matrix Profiles & S & \good & \bad & \bad & \bad & \good & \bad & \bad & B & \bad & \good$^S$ & 1 & 0 & \water & EtherNet/IP \\
		&\multirow[c]{-13.7}{*}{\rotatebox[origin=t,]{90}{\centering\textbf{Behavior Prediction}}}& SAVIOR \cite{quinonez2020savior} \litref{44} & 2020& Physical Invariants & S & \good & \bad & \bad & \bad & \good & \bad & \ok & B &\good & \bad & 2 & 1 & \transport & MAVLink \\
		\cdashlinelr{2-20}

		&& Krotofil \etal \cite{krotofil2015process} \litref{60} &2015 & Entropy Analysis & S & \good & \bad & \bad & \bad & \good & \bad & \bad & B & \bad & \ok$^T$ & 1 & 0 & \chemical &	-- \\
		&& Alippi \etal \cite{alippi2016model} & 2016 & Hidden Markov Model & S & \good & \bad & \bad & \bad & \good & \bad & \bad & B & \ok & \bad & 3 & 0 & \synthetic & -- \\ %
		&& Aggarwal \etal \cite{aggarwal2018corgids} & 2018 & Hidden Markov Model & S & \good & \bad & \bad & \bad & \good & \bad & \bad & B & \bad & \bad & 2 & 0 & \medical \vspace{.001mm} \transport & -- \\ %
		&& Hau \etal \cite{hau2019exploiting} \litref{43} & 2019& Statistics & S & \good & \bad & \bad & \bad & \good & \bad & \bad & B & \bad & \ok & 1 & 0 & \medical & -- \\
		&& NoiSense \cite{ahmed2020noisense} \litref{1} &2020 & Noise Fingerprinting & S & \good & \bad & \bad & \bad & \good & \bad & \ok & B & \bad & \good$^S_W$ & 2 & 0 & \water \hspace*{.1mm} \dist& EtherNet/IP \\
		&\multirow[c]{-6.2}{*}{\rotatebox[origin=t,]{90}{\centering\textbf{Correlations}}}& ProcessSkew \cite{ahmed2020process,ahmed2018noise} \litref{5/33} & 2020 & Noise Fingerprinting & S & \good & \bad & \bad & \bad & \good & \bad & \good & B & \bad & \good$^S$ & 1 & 0 & \water & EtherNet/IP \\ %
		\cdashlinelr{2-20}

		&& Nader \etal \cite{nader2014lp} & 2014 & SVDD, KPCA & S & \good & \bad & \bad & \bad & \good & \bad & \bad & B & \ok & \bad & 2& 0 & \gas \vspace{.001mm} \water & -- \\ %
		&& Junejo \etal \cite{junejo2016behaviour} \litref{4} &2016 & Machine-learning & S & \ok & \bad & \bad & \bad & \good & \bad & \bad & A/B & \bad & \good$^S$ & 1 & 0 & \water & EtherNet/IP \\
		&& Inoue \etal \cite{inoue2017anomaly} \litref{9} &2017 &  SVM, DNN & S & \ok & \bad & \bad & \bad & \good & \bad & \bad & B & \bad & \good$^S$ & 1 & 0 & \water & EtherNet/IP \\
		&& Chen \etal \cite{chen2018learning} \litref{51} & 2018 &  SVM & S & \bad & \bad & \bad & \bad & \good & \bad & \bad & A/B & \ok & \good$^S$ & 1 & 0 & \water & EtherNet/IP \\
		&& AADS \cite{abdelaty2019aads} \litref{47} & 2019 & DNN & S & \ok & \bad & \bad & \bad & \good & \bad & \bad & B & \bad & \good$^S$ & 1 & 3 & \water & EtherNet/IP \\
		&& Anton \etal \cite{duque2019security} \litref{6} &2019 &  OCVM, Isolation Forest & S & \bad & \bad & \bad & \bad & \good & \bad & \bad & B & \bad & \good$^S$ & 1 & 0 & \water & EtherNet/IP \\
		\multirow[c]{-40}{*}{\rotatebox[origin=t, ]{90}{\parbox{6cm}{\centering\textbf{Process State-aware Approaches (\secref{sec:iids-survey-state}) }}}} &\multirow[c]{-7.6}{*}{\rotatebox[origin=t,]{90}{\centering\textbf{Classifiers}}} & FALCON \cite{sapkota2020falcon} \litref{46} & 2020&  LSTM+ML & S & \good & \bad & \bad & \bad & \good & \bad & \bad & A/B & \bad & \good$^S$ & 1 & 2 & \water & EtherNet/IP \\

		\toprule

		\multicolumn{20}{c}{\power: Power Grid \quad \water: Water Treatment \quad \gas: Gas \quad \manufacturing: Manufacturing\quad \medical: Medical \quad \chemical: Chemical \quad \dist: Water Distribution \quad \transport: Logistics \quad \synthetic: Synthetic Setup} \\[1mm]
		\multicolumn{20}{c}{S: State, P: Packet \hspace{3em} \good: yes, \ok: partial, \bad: no \hspace{3em} A: Anomalous, B: Benign \hspace{3em} \good$^S$: SWaT~\cite{goh2016dataset}, \good$^W$: WADI~\cite{ahmed2017wadi}, \good$^T$: TEP~\cite{downs1993plant} \good$^M$: Morris~\cite{morris2015industrial}}
	\end{tabularx}
	\vspace{0.5em}
	\caption{
	Our %
	survey of 53 intrusion detection approaches confirms the heterogeneity across the industrial research landscape.
	While \acp{IIDS} operate on few information types, they are mostly developed in isolated silos and seldomly compare to existing research.
	Moreover, we observe similar detection methodologies across different protocols and domains, indicating an enormous potential for realizing protocol-independent \acp{IIDS} to transfer achievements to a broader scale of industries.
	}
	\label{tab:iids-summary}
\end{table*}

\subsubsection{Communication-based Intrusion Detection}
\label{sec:iids-survey-network}

Different streams of research propose to leverage the inherent communication patterns found across industrial domains.
We identify three categories of how regularities caused by the periodic distribution of, e.g., new sensor measurements, can be utilized for anomaly detection.
First, irregularities in %
\emph{timings} indicate potential %
malicious activity.
Secondly, the order of message \emph{sequences} also exposes repeating patterns that can be learned to identify expected normal communication. %
Finally, a class of \emph{classifier}-based approaches grade each packet as benign or anomalous based on a set of predefined features.

Our analysis shows that all three categories are used across different industrial domains.
Analyzing the type of information these %
\acp{IIDS} operate on, we observe that timing- and sequence-based approaches mostly rely on just four data types:
timing information, communicating entities, message types, and often detailed information about transmitted data.
In addition, classifier-based approaches often incorporate the exchanged values. %
Rarely do such \acp{IIDS} use further packet features, \eg packet sizes or checksums for their classification.
Still, none of the approaches take advantage of protocol-specific information, such that, in theory, they could be employed for other protocols and scenarios (for Chu~\etal~\cite{chu2019industrial}, this claim cannot be validated as their automatic feature extraction approach could potentially result in the use of arbitrary features). %

Overall, communication-based approaches are not based on the particular structure and behavior of the underlying industrial protocol.
Instead, despite \acp{IIDS} being tailored to specific protocols, they nearly exclusively use protocol-agnostic information to detect anomalies.
Hence, in theory, virtually all considered communication-based \acp{IIDS} work protocol-independent and thus show potential for deployment in a wide range of industrial domains.

However, such potential for protocol-independence remains mostly unexplored in practice.
As a rare exception, Lin~\etal evaluate their \ac{IIDS}~\cite{lin2017timing}, checking for irregular timings between similar messages, for three industrial protocols.
Still, they focus on a single domain (power grids) and thus do not examine generalizability. %
Generally, despite theoretical generalizability to other domains and protocols, communication-based \acp{IIDS} are not compared against each other or evaluated across different domains.
Consequently, researchers cannot determine which \acp{IIDS} show the best performance and should be used as a benchmark for novel proposals, improved in future research, or deployed in practice.

\subsubsection{Process State-aware Intrusion Detection}
\label{sec:iids-survey-state}

In contrast to the communication-based
approaches, process state-aware \acp{IIDS} solely base on (time-)series of industrial system's physical states.
A state comprises the combined process information of all sensors and actuators aggregated over multiple packets to assess whether a process' physical state indicates an anomaly over time.
Thus, \acp{IIDS} from this category do not rely on packet-specific information and already naturally abstract from many industrial protocol's characteristics.
To identify attacks, process state-aware approaches leverage the repetitiveness and predictability of physical processes, e.g., \acp{PLC} controlling a pump to keep a water reservoir's fill level within certain bounds~\cite{feng2019systematic}.

While process state-aware \acp{IIDS} operate on a more restricted set of information than communication-based \acp{IIDS}, their detection methodologies show more diversity, which we broadly classify along four categories.
The first category of approaches defines \emph{critical states} of an industrial system, \eg through externally provided system states by domain experts, and raises an alarm if such a state is reached.
Complementary, the second category of \acp{IIDS} attempts to \emph{predict} future states based on past observations, raising an alarm if the behavior significantly deviates from the prediction.
The third category of approaches takes a more local view by investigating \emph{correlations} between an individual or a small set of sensor readings over time, searching for indicators of anomalous activities, such as outliers from learned clusters.
Finally, as for network-based approaches, a category of approaches train \emph{classifiers} to distinguish between genuine and anomalous behavior based on process state.

The majority of approaches are evaluated on public datasets, but essentially split across merely two datasets (SWaT~\cite{goh2016dataset} and TEP~\cite{downs1993plant}), suggesting an evaluation bias.
Only Aoudi~\etal evaluate their \ac{IIDS}~\cite{aoudi2018truth} on both of these datasets.
Thus, our analysis of process state-aware approaches' evaluation methodologies again indicates that most publications neither compare their approach to related work sufficiently nor evaluate on multiple datasets.
This observation is surprising since process state-aware \acp{IIDS} are especially interesting in terms of transferability, as their underlying process representation contains information available across many industrial protocols and domains.

\subsubsection{Meta-analysis of Current IIDS Research}
\label{sec:iids-survey-meta}

Our analysis of %
53 \acp{IIDS} confirms our initial observation from \secref{sec:iids:landscape} that the \ac{IIDS} research landscape indeed suffers from \emph{heterogeneity}, and the overall progress in this research field is thus unnecessarily slowed down.
Further confirming the problem of prevalent \emph{evaluation bias}, we find that \acp{IIDS} are compared to a median of zero (average 0.38) existing \acp{IIDS}.
This phenomenon can partly be explained by the exclusive use of proprietary evaluation datasets (47\,\% of the proposals), a lack of published code or artifacts (75\,\%), and single evaluation scenarios (75\,\%).
Similarly, concerning dataset diversity, the median number of evaluated datasets is 1 (average 1.32), and 52\,\% (13 of 25) evaluating on public datasets did so exclusively on SWaT~\cite{goh2016dataset}.

Laying out a path towards protocol-independent \acp{IIDS} as a remedy for current limitations of \ac{IIDS} research (cf.~\secref{sec:iids:landscape}), we observe that \ac{IIDS} models do not depend on specific communication protocols but rather on the knowledge of which information is exchanged at which time between whom.
Furthermore, %
two-thirds (66\,\%) operate on a state of the underlying process only, completely oblivious of how and when this information is exchanged.
Thus, our survey indicates a huge potential for realizing protocol-independent \acp{IIDS}. %

\section{Proposing IPAL -- An Industrial Protocol Abstraction Layer}
\label{sec:ipal}

\begin{figure}[t]
	\centering
	\includegraphics[width=\columnwidth]{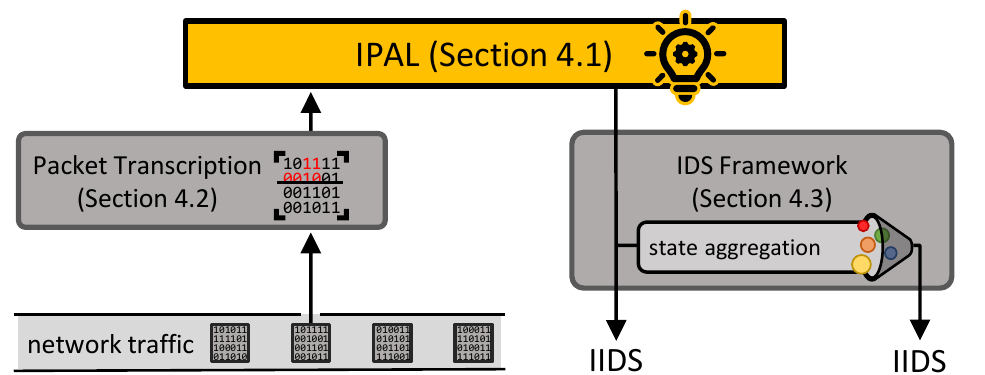}
	
	\caption{To achieve protocol independence in \ac{IIDS} research, \ac{IPAL} separates the detection methods from the underlying industrial network protocol with an abstract representation.}
	\label{fig:ipal:ids}
\end{figure}

Given the enormous potential for realizing protocol-indepen\-dent \acp{IIDS} (cf.\ \secref{sec:iids-survey}) and thus capitalize on their manifold benefits (cf.\ \secref{sec:abstract-industrial-communication:benefits}), we set out to turn this vision into reality.
To this end, we propose \ac{IPAL}, our design of an \emph{industrial protocol abstraction layer} to decouple intrusion detection from domain-specific industrial communication protocols.
As shown in \figref{fig:ipal:ids}, we %
derive \ac{IPAL} from the knowledge gathered in our survey (\secref{sec:ipal:design}).
We discuss how the landscape of industrial protocols can be transcribed into \ac{IPAL} %
(\secref{sec:ipal:ipal}) and how \acp{IIDS} implemented on top of \ac{IPAL} help transfer their potential to new scenarios %
(\secref{sec:ipal:implementation}).
Finally, we discuss the limitations of introducing an abstraction layer %
(\secref{sec:ipal:limitations}).

\subsection{Designing IPAL}
\label{sec:ipal:design}

\ac{IPAL}, our design of an \emph{abstract representation} for industrial protocols, lays the foundation to design and study protocol-independent \acp{IIDS} by unifying industrial communication.
During our comprehensive survey of \acp{IIDS} across various industries and communication protocols, we identified inherent similarities with regard to the required data~(\cf \secref{sec:iids-survey}).
Leveraging this knowledge, we deduce the information captured by \ac{IPAL}, as shown in~\tabref{tab:ipal}.
First, process state-aware \acp{IIDS}~(\cf~\secref{sec:iids-survey-state}) only operate on a very restricted set of information, encompassing only timings and information about exchanged state information.
Meanwhile, communication-based \acp{IIDS}~(\cf~\secref{sec:iids-survey-network}) operate on a broader set of information but still only rely on packets with (direct) influence on the physical process, such that other packets~(e.g., diagnostic data or TCP handshakes) can be ignored.
As detailed in the following, ten features suffice to represent industrial network packets in a common format while preserving all information required by process-aware \acp{IIDS}.

\begin{table}[t]
	\centering \footnotesize
	
	\begin{tabularx}{\columnwidth}{l@{\hskip 6mm}l}
		\toprule
		
		\textbf{IPAL}& \textbf{Description} \\ \midrule
		
		\code{id} 			& Unique identifier for \ac{IPAL} messages within a dataset \\
		\code{timestamp} 	& Time corresponding to this IPAL message \\
		\code{length} 		& Length of the industrial protocol's layer \\
		\code{malicious} 		& Label for training and evaluation (benign or anomalous) \\
		\hdashline
		\Tstrut \code{source} 		& Sender of the network packet (e.g., IP and port) \\
		\code{destination} 	& Receiver of the network packet (e.g., IP and port) \\
		\hdashline
		\Tstrut \code{message type}	& Fine-granular industrial packet's message type \\
		\code{activity} 		& Abstracted message type (e.g., request or response) \\
		\code{responds to} 	& List of related \ac{IPAL} messages this message responds to \\
		\hdashline
		\Tstrut \code{process data}	& Arbitrary number of process variables and their values \\
		\bottomrule
	\end{tabularx}
	\vspace{0.5em}
	\caption{IPAL captures ten features in an abstract representation to support current \acp{IIDS} across multiple domains.}
	\vspace{-.5em}
	\label{tab:ipal}
\end{table}

\textbf{Meta Data.}
The first observation in our survey regarding the abstraction %
is the incorporation of metadata %
across all \ac{IIDS} categories.
Thus, \ac{IPAL} includes a packet's \code{timestamp} and \code{length}. %
In addition, a unique identifier (\code{id}) is included to disambiguate packets.
Since several \acp{IIDS} require labeled training data differentiating benign and malicious packets, IPAL includes the \code{malicious} field.

\textbf{Addressing Information.}
Besides meta data, communication-based \acp{IIDS} require features to identify the \emph{communicating entities}.
Thus, \ac{IPAL} includes the \code{source} and the \code{destination} of a single packet.
These fields are arbitrary strings that, in most cases, represent an IP-port combination, but can be extended, \eg by adding Modbus's unit identifier field, to further disambiguate devices.
Likewise, \code{destination} can remain empty for broadcast protocols.

\textbf{Message Identification.}
One relevant aspect found during our survey is that communication-based \acp{IIDS} leverage %
\emph{message types} of industrial packets on different levels of abstraction~(\cf~\secref{sec:iids-survey-network}).
Some \acp{IIDS}~\cite{caselli2015modeling,caselli2015sequence} analyze the sequence of specific, reoccurring messages, and further \acp{IIDS}~\cite{lin2017timing} only differentiate between requests or responses.
Therefore, \ac{IPAL} captures a fine-granular \code{message type}, as well as a more generic \code{activity}.
Here, four \code{activities} (requests, commands, and their respective answers) suffice~\cite{wolsing2020facilitating}, while the \code{message type} closely matches the differentiation schemes used by the underlying protocols.
Lastly, the \code{responds to} field lists the \code{id}s of all \ac{IPAL} packets a given message is a response to.

\textbf{Process Data.}
The most important feature used by both types of \acp{IIDS} is the process state.
In \ac{IPAL}, the field \code{process data} collects all process variables as well as their current values as communicated within a single packet.
Depending on the transcribed packet, \code{process data} thus can range, \eg from a single temperature to an array of sensor readings and actuator commands.
As process state-aware \acp{IIDS} demand a snapshot of the entire system's state, we support state aggregation over several \ac{IPAL} messages to realize a similar extraction as, \eg through active polling~\cite{PLC4X}.

A few rare cases exist where additional features (\eg checksum validity tests~\cite{chu2019industrial}) are included in the decision-making process of otherwise process-aware \acp{IIDS}.
These checks are, however, already covered by traditional IDS~(\eg~Zeek~\cite{Zeek}) and do not contribute to information about the physical process monitored by process-aware \acp{IIDS}, and can thus be skipped to avoid redundant work.
Consequently, the ten features of IPAL suffice to preserve all information required by the process-aware \acp{IIDS} surveyed in~\tabref{tab:iids-summary}.

\subsection{Transcribing Industrial Protocols}
\label{sec:ipal:ipal}

IPAL, our abstract representation of industrial network traffic, pro\-mises to decouple process-aware intrusion detection from underlying industrial protocols.
Yet, to realize protocol-independent~\acp{IIDS}, real-world network traffic must be transferred into this abstract representation.
As shown in~\figref{fig:ipal:ids}, we designed a \emph{transcriber}~\cite{GitIPALTranscriber} that automates this conversion.
While integrating specific industrial protocols into \ac{IPAL} requires a deep protocol understanding, this work needs to be performed only once.

We have already incorporated nine industrial protocols (Modbus, EtherNet/IP, MQTT, IEC-104, DNP3, S7, IEC 61850-GOOSE, IEC 61162-450, and NMEA~0183) used in popular datasets.
Thus, future \ac{IIDS} research can focus on modeling industrial processes, while a broad set of datasets from different industrial scenarios is readily available for evaluation through \ac{IPAL}~\cite{GitIPALDatasets}.
In the following, we discuss challenges faced while implementing our transcriber. %

Most features of \ac{IPAL} can be directly extracted from network packets.
However, certain protocols require special attention.
E.g., in ModbusTCP, multiple sub-devices may operate behind a single TCP connection.
Thus, \code{source} and \code{destination} need to include not only the IP and port but also the device's unit identifier.
Also, while Modbus's function code directly corresponds to the \code{message type}, the \code{activity} has to be derived from the traffic direction, as function codes for requests and responses are identical.

A more challenging aspect is the extraction of the actual process values from industrial communication because of their encoding.
To this end, transcribers can contain custom rules to extract and post-process process values from observed communication, similar to  StreamPipes~\cite{zehnder2020streampipes}.
Such rules are small code snippets that allow, e.g., to interpret two 16-bit registers (the biggest data type in Modbus) as a single 32-bit \,float. %
Thus, we can integrate the necessary flexibility to define the interpretation of potentially multiple related variables and even directly annotate the interpreted value with a more descriptive and human-readable name.

\subsection{Developing IIDSs with IPAL}
\label{sec:ipal:implementation}

To leverage its benefits, we provide a framework to implement protocol-independent \acp{IIDS} on top of \ac{IPAL}~\cite{GitIPALIIDS}.
We observed in our survey~(\cf~\secref{sec:iids-survey}) that \acp{IIDS} are either based on communication traffic or on the global process state.
As shown in \figref{fig:ipal:ids}, for communication-based \acp{IIDS}, our framework passes each packet to the \ac{IIDS} after abstracting it into an \ac{IPAL} message.

For process state-aware \acp{IIDS}, we need to aggregate multiple packets over time to obtain the current state of the entire process.
Thus, our framework simply caches the process values included within the most recent \ac{IPAL} messages.
We considered different state aggregation methods, but simply outputting the most recent process values in regular intervals~(\eg~each second) yields the best results and allows us to convert EtherNet/IP traffic included in a newer run of the SWaT testbed~(without attacks) to a state that closely matches the pre-processed state information provided by the dataset authors.
Furthermore, as some common datasets %
(\eg~SWaT~\cite{goh2016dataset}) already provide a state representation, our framework optionally allows restoring these pre-computed states directly. %

Our framework enables the realization of communication-based and process state-aware \acp{IIDS} on \ac{IPAL} messages.
\ac{IPAL} already supports nine industrial protocols from various domains. %
Thus, \acp{IIDS} developed for or adjusted to \ac{IPAL} can be evaluated for various scenarios based on current and future datasets, accelerating research on \acp{IIDS} and moving them closer to deployment~(\cf~\secref{sec:abstract-industrial-communication:benefits}).

\subsection{Limitations of Protocol Abstraction}
\label{sec:ipal:limitations}

Any form of abstraction involves the risk of introducing an information loss, as crucial features may be discarded.
In our case, \acp{IIDS} taking advantage of a feature not covered by our abstraction would likely not be compatible with IPAL.
However, considering that IPAL's features are derived from an extensive survey of 53 scientific publications (cf.\ \secref{sec:iids-survey}), we expect the number of currently developed \acp{IIDS} that cannot be realized on top of IPAL to be low.

Still, our choice of features captured in IPAL could potentially narrow the scope of future research.
However, for dominantly used datasets, we already observe a restriction in available information to pre-processed data, \eg only providing timing and process-state information (\cf SWaT in \tabref{tab:iids-summary}), presumably to facilitate easy utilization of the dataset.
Contrary to intuition, the adoption of IPAL could %
even lead to the availability of more features %
by encouraging the release of unprocessed packet captures, as IPAL provides the tooling to easily process raw data according to researchers' needs.
In this vein, the IPAL framework is publicly available~\cite{GitIPALTranscriber,GitIPALIIDS} such that the extracted feature set can be extended in the future if deemed necessary.
A similar trend has been observed for Netflow~\cite{RFC3954}, whose abstraction revolutionized network monitoring and introduced newer versions as the need for more features arose. %

From a different perspective, protocol abstraction comes with the potential risk of introducing transcription or abstraction inaccuracies affecting the final detection performance. %
Yet, as we show in the following (\secref{sec:reproducibility}), such inaccuracies are generally not introduced by IPAL, and the \ac{IIDS} performance is independent of the use of our abstraction layer.
Consequently, the value of IPAL homogenizing a split-up research landscape far outweighs the potential drawbacks arising from introducing an abstraction layer (so early) for the growing industrial intrusion detection research field.

\section{Practical Applicability of IPAL} %
\label{sec:reproducibility}

\begin{table}[t]
	\centering \footnotesize

	\begin{tabularx}{\columnwidth}{l@{\hskip 6mm}l@{\hskip 6mm}ccc}
		\toprule
		\textbf{Category} & \textbf{IIDS} & \textbf{Code} & \textbf{Dataset}& \textbf{Result} \\

		\midrule

		Communication & Inter arrival time~\cite{lin2017timing} & \bad & \ok & \checkmark \\
		& DTMC~\cite{ferling2018intrusion} & \good & $\RIGHTcircle$ & (\checkmark) \\
		& RF~\cite{perez2018machine} & \good & \good & \checkmark \\
		& SVM~\cite{perez2018machine} & \good & \good & \checkmark \\
		& BLSTM~\cite{perez2018machine} & \good & \good & \checkmark \\ \midrule

		Process state & PASAD~\cite{aoudi2018truth} & \good & \good & \checkmark \\
		& Seq2Seq-NN~\cite{kim2019anomaly} & \good & \good & \checkmark \\
		& TABOR~\cite{lin2018tabor} & \good & \good & \checkmark \\
		\bottomrule
		\multicolumn{5}{c}{\good: available \ok: dataset $\RIGHTcircle$: attacks \bad: no \quad (\checkmark) no comparison to original paper}
	\end{tabularx}
	\vspace{0.5em}
	\caption{We prove the applicability and correctness of \ac{IPAL} by successfully reproducing eight \acp{IIDS} on top of \ac{IPAL}. These cover all categories from the survey and were selected by the availability of code and evaluation datasets.}
	\label{tab:reproduction-ids-selection}
\end{table}

We derived \ac{IPAL} from the requirements of existing \acp{IIDS} and discussed its theoretical applicability for industrial network protocols.
Now, we complement this viewpoint by showing \ac{IPAL}'s \emph{practical applicability and correctness} by performing a reproducibility study of eight \ac{IIDS} approaches from previous work and reproducing their evaluation results on top of \ac{IPAL}.
Thereby, we also contribute to reproducing scientific research results as an independent party.
While deemed extremely important \cite{bajpai2019reproducibility}, this is often a challenging and tedious task with little reward and thus rarely performed~\cite{erba2020no,GitPyPASAD}.

\textbf{IIDS selection criteria.}
To conduct a comprehensive reproducibility study, we set out to reproduce at least three representative approaches for each of the two major categories in our survey (cf. %
\tabref{tab:iids-summary}), covering all features provided by \ac{IPAL}.
Since we re-evaluate existing \acp{IIDS}, we rely on the availability of the original evaluation dataset or tools for attack generation.
Also, as re-implementing \acp{IIDS} from scratch can be challenging, we focus on open source approaches and only implement \acp{IIDS} ourselves if necessary.
Thus, our selection was driven by the availability of code or datasets.

Note that we reproduce evaluation results and thus do not invent new attack methodologies.
This implies that the reproducibility results are not necessarily comparable, %
as the authors used fairly different evaluation procedures. %
Still, we will show in \secref{sec:casestudies} how IPAL enables conducting a side-by-side evaluation of different \acp{IIDS}.

In the following, we briefly argue why we selected each \ac{IIDS}, summarize its core idea, provide re-implementation and evaluation details, and compare our results %
to the original publication.
\tabref{tab:reproduction-ids-selection} summarizes our \ac{IIDS} selection %
and reproduction results.

\subsection{Inter-arrival Time (Communication)}

\begin{table}[t]
	\centering \footnotesize

	\begin{tabular}{lcl@{\hskip 1mm}rr rr}
		\toprule
		\multirow{2}{*}{\rotatebox[origin=c]{90}{\textbf{Attack}}} &  \multirow{2}{*}{\vspace{-1.7mm} \textbf{Model}} & \multirow{2}{*}{\vspace{-1.7mm} \textbf{Metric}} & \multicolumn{2}{c}{\textbf{Original}} & \multicolumn{2}{c}{\textbf{IPAL}} \\
		\cmidrule(lr){4-5} \cmidrule(lr){6-7}
		&&& \multicolumn{1}{c}{Request } & \multicolumn{1}{c}{Response } & \multicolumn{1}{c}{Request} & \multicolumn{1}{c}{Response} \\
		\midrule

		\multirow{5}{*}{\rotatebox[origin=c]{90}{Flooding}}
		& \multirow{2}{*}{\vspace{-0.3mm} Mean}
		& TPR & 99.90 & 99.90 & 99.98 & 99.97 \\
		&& FPR & 0.01 & 0.20 & 0.01 & 0.01 \\
		\cmidrule(lr){2-7}
		& \multirow{2}{*}{\vspace{-0.3mm} Range}
		& TPR & 59.10 & 56.40 & 68.21 & 65.48 \\
		&& FPR & 0.80 & 1.10 & 0.98 & 1.20 \\
		\cmidrule(lr){2-7}
		&& ODR & 100.00 & 100.00 & 100.00 & 100.00 \\
		\midrule

		\multirow{5}{*}{\rotatebox[origin=c]{90}{Injection}}
		& \multirow{2}{*}{\vspace{-0.3mm} Mean}
		& TPR & 96.20 & 96.60 & 98.95 & 98.95 \\
		&& FPR & 0.01 & 0.20 & 0.01 & 0.01 \\
		\cmidrule(lr){2-7}
		& \multirow{2}{*}{\vspace{-0.3mm} Range}
		& TPR & 100.00 & 99.50 & 99.47 & 99.47 \\
		&& FPR & 0.80 & 1.10 & 1.01 & 1.21 \\
		\cmidrule(lr){2-7}
		&& ODR & 100.00 & 100.00 & 97.37 & 97.37 \\
		\midrule

		\multirow{5}{*}{\rotatebox[origin=c]{90}{Prediction}}
		& \multirow{2}{*}{\vspace{-0.3mm} Mean}
		& TPR & 0.10 & 0.20 & 4.60 & 4.92 \\
		&& FPR & 0.00 & 0.20 & 0.01 & 0.01 \\
		\cmidrule(lr){2-7}
		& \multirow{2}{*}{\vspace{-0.3mm} Range}
		& TPR & 90.60 & 91.00 & 93.69 & 93.59 \\
		&& FPR & 0.80 & 1.10 & 0.97 & 1.28 \\
		\cmidrule(lr){2-7}
		&& ODR & 99.80 & 99.50 & 95.21 & 95.23 \\
		\bottomrule

	\end{tabular} \\
	TPR: True Positive Rate, FPR: False Positive Rate, ODR: Overall Detection Rate (in \%)
	\vspace{0.5em}
    \caption{Our re-implemented IAT \ac{IIDS} %
	closely resembles the original detection rates (cf. Tab. 2 in~\cite{lin2017timing}). Slight deviations result from minor differences in the attack generation.}
	\label{tab:reproduction-lin17}
\end{table}

\acp{IIDS} from the communication-based category utilize data on a per-packet basis and consider features such as communicating entities and message types.
An ideal representative for this category is the inter-arrival time (IAT) approach by Lin et al.~\cite{lin2017timing}, which utilizes periodic traffic patterns in industrial protocols such as Modbus, S7, and IEC-104 to detect, e.g., packet-injection attacks.
To this end, the approach measures the \emph{mean} inter-arrival time, i.e., the elapsed time between two packets of the same type (e.g., requests and responses), as well as the maximum temporal deviation (\emph{range}) between packets with the same content and checks for timing violations.

As no implementation of the inter-arrival time approach was available, we re-implemented it based on the paper~\cite{lin2017timing}.
To reproduce the results, we use the only publicly available dataset from the original work~\cite{lin2017timing}, which evaluates S7 traffic~\cite{s7dataset}.
As only the attack-free dataset was available, we performed an identical 1/10 train-test dataset split and injected ``malicious'' packets according to the authors' description.
Our reproduced results on top of \ac{IPAL} (cf. \tabref{tab:reproduction-lin17}) closely match those of the original paper.
We attribute the remaining differences to randomness %
and minor implementation uncertainties \wrt to attack generation.
Overall, %
\ac{IPAL} provides everything needed to re-implement and reproduce this \ac{IIDS}.

\subsection{DTMC (Communication)}

As an additional communication-based \ac{IIDS}, we consider an approach relying on \acfp{DTMC} modeling the sequence of network packets for a single connection~\cite{ferling2018intrusion}.
This approach leverages a core property of industrial networks:
Devices such as \acp{PLC} perform several communication steps in series, e.g., reading values from remote devices and adjusting setpoints later on, resulting in periodic packet sequences repeating in similar order.

We adapted the \ac{DTMC} approach's publicly available source code~\cite{Gitintravis} to operate on top of \ac{IPAL}.
While a full reproducibility study is infeasible without the original dataset, we can still validate whether the original implementation and the version on top of \ac{IPAL} produce identical results when presented with the same input.
To this end, we generated our own IEC-104 network trace using a simulation framework for power distribution grids~\cite{henze2020poster} and the attack tool by the authors to add equivalent attacks to our trace~\cite{GitmanipulateTraces}.
We performed the same procedure as in the original paper~\cite{ferling2018intrusion}, i.e., a 50/50 train/test split, applying the attack tool ten times for each attack type and counting the number of state and transition violations compared to the trained \ac{DTMC}.
While not comparable to the original evaluation, \ac{IPAL} correctly preserves all information resulting in identical attack coverage compared to the original \ac{DTMC} implementation, as shown in~\tabref{tab:reproduction-dtmc}.

\begin{table}[t]
	\centering \footnotesize

	\begin{tabularx}{\columnwidth}{c@{\hskip 2mm} c@{\hskip 2mm} c@{\hskip 2mm}c@{\hskip 2mm}c c@{\hskip 2mm}c@{\hskip 2mm}c c@{\hskip 2mm}c@{\hskip 2mm}c} \toprule
		\multirow{2}{1.2cm}{ \centering \textbf{Reduction Type}} & \multirow{2}{1.1cm}{ \centering \textbf{Anomaly Type}} & \multicolumn{3}{c}{\textbf{\textbf{Copy [\%]}}} & \multicolumn{3}{c}{\textbf{Remove  [\%]}} & \multicolumn{3}{c}{\textbf{Swap  [\%]}} \\
		\cmidrule(lr){3-5} \cmidrule(lr){6-8} \cmidrule(lr){9-11}
		 &  & $0.1$ & $1$ & $10$ & $0.1$ & $1$ & $10$ & $0.1$ & $1$ & $10$ \\ \toprule
		\multirow[c]{2}{*}{none} & state & 0 & 0 & 0 & 0 & 0 & 0 & 0 & 0 & 0 \\
			& transition & 0 & 15.5 & 38.5 & 0 & 2 & 14 & 0 & 20.5 & 46.6 \\ \midrule
		\multirow[c]{2}{*}{overlapping} & state & 0 & 0 & 0 & 0 & 0 & 0 & 0 & 0 & 0 \\
			& transition & 0 & 8.5 & 19 & 0 & 1.5 & 6.5 & 0 & 11.5 & 22 \\ \midrule
		\multirow[c]{2}{*}{all} & state & 1 & 1 & 1 & 1 & 1 & 1 & 1 & 1 & 1 \\
			& transition & 0 & 9.5 & 22 & 0 & 1.5 & 6.5 & 0 & 13 & 27 \\ \bottomrule
	\end{tabularx}
	\vspace{0.5em}
	\caption{Adapting the \ac{DTMC} approach to \ac{IPAL} yields identical detection results in the number of state and transition validations. %
	We do not show the identical values twice here.}
	\label{tab:reproduction-dtmc}
\end{table}

\subsection{Classifiers (Communication)}

Complementing the IAT and the \ac{DTMC} approach, \acp{IIDS} using machine-learning classifiers may operate on additional information such as the packet length.
Furthermore, they additionally require traces of past attacks for training.
Thus we selected the work of Perez et al.~\cite{perez2018machine} as a representative for this category. %
In their work, they compare three different machine learning algorithms against each other (\acf{RF}, \acf{SVM}, and \acf{BLSTM}).

\begin{table}[t]
	\centering \footnotesize

	\begin{tabularx}{.75\columnwidth}{lcc@{\hskip 8mm}c}
		\toprule
		& \textbf{Paper}~\cite{perez2018machine} & \textbf{Original}~\cite{GitML} & \textbf{IPAL} \\
		\midrule

		SVM		& 94.36\,\% & 94.36\,\%  & 96.10\,\% \\
		RF 		& 99.58\,\% & 99.52\,\%  & 99.73\,\% \\
		BLSTM 	& 98.40\,\% & 98.12\,\%  & 96.78\,\% \\
		\bottomrule
	\end{tabularx}
	\vspace{0.5em}
	\caption{Measuring the accuracy of the public source code (Original) resembles the original paper results. Also, on top of IPAL, we achieve equivalent accuracy most of the time.}
	\label{tab:reproduction:dpi-ml}
\end{table}

For our reproducibility study, we adapted the implementations of the machine-learning classifiers published by the authors~\cite{GitML} to operate on top of \ac{IPAL}.
Furthermore, we use the same evaluation dataset~\cite{morris2015industrial}, hyper-parameters, and pre-processing steps as in the original paper~\cite{perez2018machine}.
In \tabref{tab:reproduction:dpi-ml}, we compare the original results of the paper with our replication of the original evaluation as well as the same evaluation on top of \ac{IPAL}.
All three approaches perform equivalently, and the minor deviations can be inferred from randomly selecting a different train and test dataset.
Thus, for these three approaches and the previous two communication-based \acp{IIDS}, we showed that \ac{IPAL} correctly preserves all relevant information to achieve equivalent attack detection performance.

\subsection{PASAD (Process state)}

In contrast to the thus far considered \acp{IIDS}, \acf{PASAD}~\cite{aoudi2018truth} detects structural changes in data series and thus represents the process state-aware category.
\ac{PASAD}'s core idea is that legit data series span a vector space, and the distance to the mean of all training vectors indicates an anomaly.
An alarm is raised if the distance (departure score) exceeds the maximum observed distance during training (threshold).

\begin{figure}[t]
	\centering
	\includegraphics{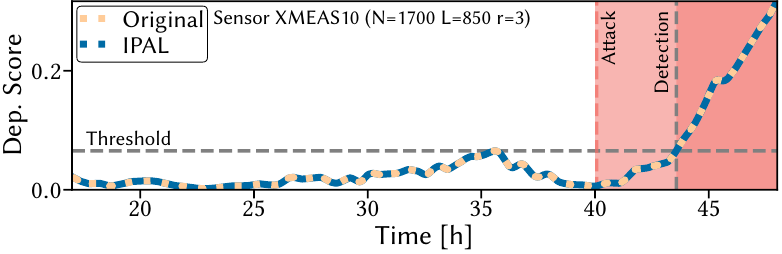}

	\vspace{-.5em}
	\caption{\ac{PASAD}'s original departure scores (yellow, cf. Fig. 4 in \cite{aoudi2018truth}) match ours %
	(blue). %
	Both %
	have identical thresholds of $\sim0.065$ and detect the attack simultaneously after $\sim3.58h$.}
	\label{fig:reproduction-pasad}
\end{figure}

Two implementations of \ac{PASAD} are available, the authors' version in Matlab~\cite{GitPASAD} and a re-implementation in Python~\cite{GitPyPASAD}.
We realized \ac{PASAD} for \ac{IPAL} based on the Python version and used the original Matlab implementation to generate reference data.
PASAD was trained and evaluated visually on output logs of the TEP~\cite{downs1993plant}, parts of the SWaT datasets~\cite{goh2016dataset}, and private data of a real water treatment facility.
We verify our implementation on the public TEP datasets and exemplary discuss one example in \figref{fig:reproduction-pasad}.
The results obtained on top of \ac{IPAL} are identical to the original implementation~\cite{GitPASAD}, and both detect the attack simultaneously.
By reproducing these results, we show that \ac{IPAL} is able to preserve all information required by a process state-aware \acp{IIDS} such as PASAD.

\subsection{Seq2Seq-NN (Process state)}

Extending on the idea of PASAD, the approach by Kim et al.~\cite{kim2019anomaly} requires knowledge about \emph{multiple} actuators and sensors to predict the expected future states of an industrial system.
More specifically, sequence-to-sequence Neural Networks (Seq2Seq-NN), known from language translation, are provided with the recent history of system states (e.g., the last $99$ seconds) and predict the next state.
Then, the distance between the predicted and actual state is calculated, and the error aggregated over time, resulting in an anomaly score.
If the anomaly score exceeds a threshold, the \ac{IDS} raises an alarm.

\begin{figure}[t]
	\centering
	\includegraphics{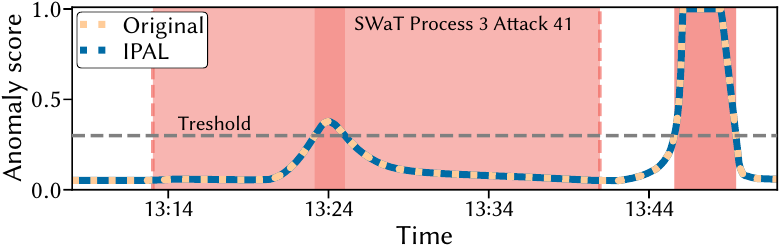}

	\vspace{-.5em}
	\caption{Seq2Seq-NN's original anomaly scores (yellow, cf. Fig.~3 in \cite{kim2019anomaly}) match our reproduced implementation (blue). %
	Both implementations emit alarms simultaneously.}
	\label{fig:reproduction-nn}
\end{figure}

For our Seq2Seq-NN implementation on top of \ac{IPAL}, we adapted its publicly available version~\cite{GitSeq2SeqNN}, extracted the original anomaly scores as a comparison baseline, and used the authors' pre-trained models for the evaluation on \ac{IPAL}.
We exemplarily compare the original implementation and the one on top of \ac{IPAL} \wrt their ability to detect an attack on stage $3$ of SWaT~\cite{goh2016dataset} in \figref{fig:reproduction-nn}.
There the anomaly score is identical, and the attack is detected simultaneously, showing \ac{IPAL} correctly providing all information for Seq2Seq-NN.

\subsection{TABOR (Process state)}

\acused{TABOR}

Complementing \ac{PASAD} and Seq2Seq-NN, \ac{TABOR} focuses on one sensor and models the impact of actuators on this sensor~\cite{lin2018tabor}.
To this end, TABOR combines three distinct detection approaches into one \ac{IIDS}: %
An out-of-alphabet check alerts if unseen data or data out of regular boundaries is observed, a timed automaton captures the sequence and duration of linear sensor value segments to enable anomaly detection, and a Bayesian network couples sensor segments with actuator states to identify unexpected combinations.

\begin{table}[t]
	\centering \footnotesize
	
	\begin{tabularx}{\columnwidth}{l X r@{\hskip 3mm}r@{\hskip 3mm}r@{\hskip 3mm}r X r@{\hskip 3mm}r@{\hskip 3mm}r@{\hskip 3mm}r} \hline
		\toprule
		\multirow{2}{*}{\vspace{-1.7mm} \textbf{Model  1}} &  & \multicolumn{4}{c}{\textbf{Original}} && \multicolumn{4}{c}{\textbf{IPAL}} \\
		\cmidrule(lr){3-6} \cmidrule(lr){8-11}
		& &\centering FP &\centering TP & \centering CP & \centering PS & & \centering FP &\centering TP & \centering CP & \multicolumn{1}{c}{PS} \\ \midrule
		
		TABOR &	& 0 & 9 & 7.85 & 1889 	&& 0 & 9 & 7.09 & 1765 \\
		TA 		&& 7 &26 &87.48 &189516 	&& 7 &26 &86.3 & 188829  \\
		BN 		&& 0 & 5 &1.95 & 629 		&& 0 & 5 &1.27 & 504  \\
		OOA 	&	& 0 & 5 &5.90 & 1260 	&& 0 & 5 &5.82 & 1261  \\
		\bottomrule
	\end{tabularx}
	\vspace{0.5em}
	\caption{\ac{TABOR}, on top of \ac{IPAL}, detects the identical nine attack scenarios (TP) with slightly different coverage percentage (CP) and penalty scores (PS) (cf. Tab. 2 in~\cite{lin2018tabor}).}
	\vspace{-0.5em}
	\label{tab:reproduction-TABOR}
\end{table}

We were able to retrieve the implementation of \ac{TABOR} from the authors for our re-implementation on top of \ac{IPAL}.
\ac{TABOR} was evaluated on the SWaT dataset~\cite{goh2016dataset}, which was divided into smaller logically coherent units (models).
Here, we concentrate on reproducing model 1 as its intermediate results in the original paper~\cite{lin2018tabor} ease verifying the correctness of our implementation.
We achieved equivalent results for our version of \ac{TABOR} on top of \ac{IPAL} (cf. \tabref{tab:reproduction-TABOR}). %
Our re-implementation scores are identical in true positives (TP) and false positives and only deviates marginally within the coverage percentage (CP) and penalty scores (PS), probably due to minor temporal deviations in the alarms.
We confirmed manually that the alerts overlap between the original and our implementation.
Concluding, \ac{IPAL} provides the necessary information to realize even complex, combined process state-aware \acp{IIDS}.

\subsection{Summary and Lessons Learned}

To showcase the practical applicability and correctness of \ac{IPAL}, we performed a reproducibility study of eight \acp{IIDS} covering the two \ac{IIDS} categories with at least three representatives (cf.~\tabref{tab:iids-summary}).
By (re-)implementing a diverse set of \acp{IIDS} on top of \ac{IPAL}, we have shown that \ac{IPAL} provides all information required by various types %
of \acp{IIDS}.
By reproducing previous work's results, we have shown that \ac{IPAL} operates correctly and neither reduces functionality nor detection rates (\cf \secref{sec:ipal:limitations}).
Consequently, \ac{IPAL} provides a reliable foundation for protocol-independent \acp{IIDS}, thus capitalizing on the benefits of an abstract representation (cf. \secref{sec:abstract-industrial-communication:benefits}).

As indicated in \tabref{tab:reproduction-ids-selection}, successfully reproducing others' research requires the availability of the original implementation.
Only in rare cases does an evaluation dataset suffices.
While we were able to successfully reproduce seven \ac{IIDS} approaches (and even achieve fully identical results for two of them), we could not reproduce the evaluation results of the DTMC approach~\cite{ferling2018intrusion} since the evaluation dataset was not available to us.
Still, as the original source code and attack tool were available, we could use another dataset to show that our re-implementation on top of \ac{IPAL} produces exactly the same results as the original implementation.

While detection performance is of utmost interest, %
adding abstraction layers, such as \ac{IPAL}, introduces overheads.
For a fast \ac{IIDS} such as DTMC, we exemplary compared the execution time against the original implementation.
\ac{IPAL} yields an overhead of just $10.1\%$ for the benefits of increased applicability.
This overhead is expected to shrink for more complex \acp{IIDS}. %

Overall, the considered eight \acp{IIDS} %
not only rely on the requirements derived from our survey (\secref{sec:ipal}) but also cover four distinct protocols (Modbus, EtherNet/IP, S7, and IEC-104) from four datasets~\cite{goh2016dataset,downs1993plant,morris2015industrial,s7dataset}.
Besides contributing to the important task of reproducing scientific research results as an independent party~\cite{bajpai2019reproducibility,erba2020no}, we lay the foundation to ease the \emph{protocol-independent} development, testing, and evaluation of \acp{IIDS} by providing a rich foundation for broad evaluations. %

\section{IIDS Generalizability Study} %
\label{sec:casestudies}

Our reproducibility study showed that existing \acp{IIDS} can be implemented on top of \ac{IPAL} without loss of functionality or detection capabilities (cf. \secref{sec:reproducibility}). %
What remains to be shown is that \ac{IPAL} is actually useful, i.e., it addresses the pressing problems of current \ac{IIDS} research (\secref{sec:iids:landscape}).
More precisely, we demonstrate how \ac{IPAL} helps to transfer an \ac{IIDS} to a new industrial protocol, how \acp{IIDS} generalize to new scenarios, how existing \acp{IIDS} compare against each other, and how distinctly communication-based and process state-aware \acp{IIDS} detect different attack types.
To this end, we perform three case studies (i) for communication-based \acp{IIDS} (\secref{sec:casestudies:comm}), (ii) for process state-aware \acp{IIDS} (\secref{sec:casestudies:control}), and (iii) combining both directions of intrusion detection (\secref{sec:casestudies:combination}).

\subsection{Generalizing Communication-based IIDSs}
\label{sec:casestudies:comm}

The first case study demonstrates \ac{IPAL}'s ability to transfer commu\-nication-based \acp{IIDS} to new and realistic scenarios.
Therefore, we study the two approaches IAT~\cite{lin2017timing} and \ac{DTMC}~\cite{ferling2018intrusion} from our reproduction study (\secref{sec:reproducibility}).
Both were previously not compared against each other and were evaluated on artificial attacks retroactively integrated into network traces.
Thus, their effectiveness in realistic scenarios is currently unknown.

To this end, we utilize novel datasets from realistic industrial systems not yet covered by both \acp{IIDS}.
The first dataset resembles a \ac{DoS} attack against a power grid communicating with the IEC 61850-GOOSE protocol~\cite{goose_dataset}.
In the same setting, more advanced attacks are performed based on a combination of message modification, injection, and replaying, comprising our second dataset~\cite{goose_dataset}.
Finally, we look at a \ac{MitM} attack dataset based on ARP poisoning to manipulate the data within Modbus packets~\cite{modbus_dataset}.
Note that these datasets already contained these attacks and were specifically designed for evaluating the performance of security solutions.
Our transcriber (cf.~\secref{sec:ipal:ipal}) transparently translates the GOOSE and Modbus network traces into \ac{IPAL}, obviating the need to modify the existing \ac{IIDS} implementations.
\tabref{tab:case-study-1} summarizes the results on these three new datasets.

\begin{table}
	\centering \footnotesize

	\begin{tabularx}{\columnwidth}{l@{\hskip 4mm} c@{\hskip 3mm}c@{\hskip 3mm}c c@{\hskip 3mm}c@{\hskip 3mm}c c@{\hskip 3mm}c@{\hskip 3mm}cc }
		\toprule
		\multicolumn{1}{c}{\textbf{Scenario}} & \multicolumn{3}{c}{\textbf{DoS~\cite{goose_dataset}}} & \multicolumn{3}{c}{\textbf{Advanced~\cite{goose_dataset}}} & \multicolumn{3}{c}{\textbf{MitM~\cite{modbus_dataset}}} \\
		\cmidrule(lr){2-4} \cmidrule(lr){5-7} \cmidrule(lr){8-10}
		\multicolumn{1}{c}{Metrics}     & prec.  & rec.  & f1 &  prec.  & rec. & f1 & prec.  & rec.  & f1  \\
		\midrule
		IAT (mean) [\%] & \textbf{99} & \textbf{100} & \textbf{99} & 0 & 0 & - & \textbf{100} & \textbf{98} & \textbf{99}\\
		IAT (range) [\%]& \textbf{90} & \textbf{100} & \textbf{95} & 0 & 62 & 1 & 99 & 31 & 47\\
		DTMC [\%] & \textbf{100} & \textbf{100} & \textbf{100} &\textbf{100} & \textbf{95} & \textbf{97} & 0 & 0 & -\\
		\bottomrule
	\end{tabularx}
	\vspace{0.5em}
	\caption{\ac{IPAL} enables the comparison of communication-based \acp{IIDS} in realistic scenarios. We see that each \ac{IIDS} detects certain attacks well in practice, while other attacks that should be detectable in theory are not caught.}
	\vspace{-0.5em}
	\label{tab:case-study-1}
\end{table}

\begin{figure*}[h!t]
	\centering
	\includegraphics[width=\textwidth]{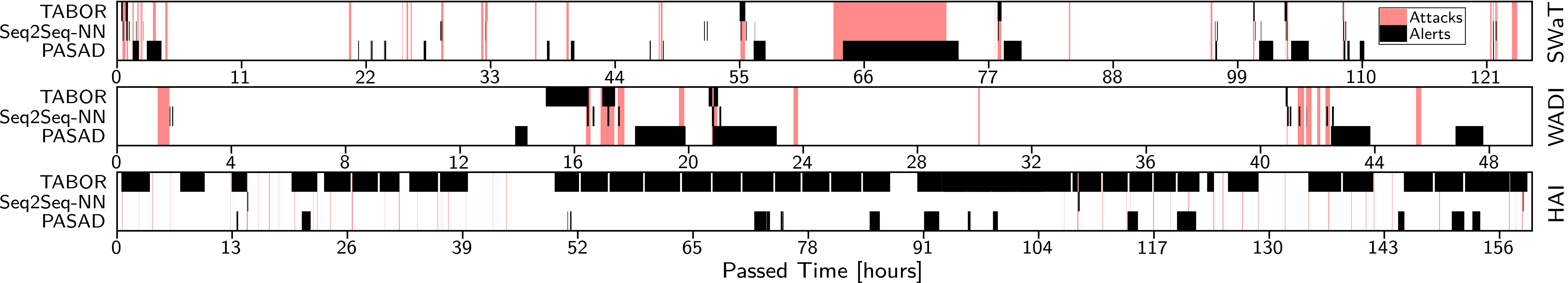}
	
	\caption{\ac{IPAL} allows the side-by-side comparison of three process state-aware \acp{IIDS} on industrial datasets and simultaneously enables to understand how \acp{IIDS} generalize to similar (WADI) and new domains (HAI).}
	\label{fig:casestudies:comparison}
\end{figure*}

\textbf{DoS.}
All \acp{IIDS} detect the packets injected by the DoS attacks to a high degree.
For the IAT's mean model, these results are expected, as this attack is similar to the artificial flooding attack evaluated in the original publication~\cite{lin2017timing}.
Surprisingly, even the IAT's range model achieves good detection rates, even though it performed worse in the artificial scenario (cf. \tabref{tab:reproduction-lin17}).
However, the DTMC achieved the best performance, classifying all packets correctly despite no similar attack being discussed in the original paper outperforming an \ac{IIDS} specifically designed to detect such attacks.

\textbf{Advanced.}
The second scenario considers advanced data manipulation attacks against a power grid similar to the artificial attacks evaluated in both original publications~\cite{lin2017timing,ferling2018intrusion}.
Still, neither of the IAT models can sufficiently detect these attacks, despite performing well in the artificial attacks originally evaluated.
This behavior can be explained by the low number of data manipulations and variable retransmission timings of the GOOSE protocol, leading to a larger variety of benign inter-arrival times.
On the other hand, DTMC achieves great detection rates, confirming the results from the original publication in a new industrial domain and protocol.

\textbf{MitM.}
Finally, we consider a MitM attack manipulating data received by industrial controllers~\cite{modbus_dataset}.
While such an attack is a realistic threat in practice, it has neither been considered by IAT nor by DTMC thus far.
Still, %
the IAT's mean model performs especially well in this scenario %
without issuing any false alarm.
Contrary, DTMC is not able to detect a single malicious packet since this attack does not change the sequence of observed messages.

These results indicate that artificial attacks, retroactively integrated into a dataset~\cite{lin2017timing,ferling2018intrusion}, %
do not necessarily carry over to more realistic attack scenarios, even when considering a similar attack behavior.
Yet, these new attack types %
can be detected well in some instances.
This observation highlights that \acp{IIDS} have to be studied more extensively to understand which attacks they can detect and which not.
Additionally, our case study highlights that one \ac{IIDS} does not necessarily have to be superior %
in every attack scenario. %
Thus, cooperative \acp{IIDS}, \ie several detecting specific attack classes, may improve overall performance (cf.~\secref{sec:casestudies:combination}). %
Finally, we showed the adaption from one domain or protocol to another scenario, confirming the enormous potential for protocol-independent and domain-agnostic \acp{IIDS} as previously identified in \secref{sec:abstract-industrial-communication:benefits}.

\subsection{Generalizing Process state-aware IIDSs}
\label{sec:casestudies:control}

Our second case study focuses on %
process state-aware \acp{IIDS}.
Since the generalizability of \acp{IIDS} has not been evaluated sufficiently before (cf. \secref{sec:iids:landscape}), we show how \ac{IPAL} eases the adaption of \acp{IIDS} to new domains and thus understand how \acp{IIDS} transfer to new scenarios.
Furthermore, as Urbina et al. \cite{giraldo2018survey} already observed, approaches are insufficiently compared in the literature. %
E.g., PASAD is evaluated only visually~\cite{aoudi2018truth}, while TABOR defines its own metrics~\cite{lin2018tabor}, and Seq2Seq-NN counts an attack as detected even $15$ minutes after the attack ended~\cite{kim2019anomaly} (\cf \secref{sec:reproducibility}).
Therefore, we evaluate these approaches against each other in a detailed comparison.

From our reproducibility study, we have three %
process state-aware \acp{IIDS} (TABOR~\cite{lin2018tabor}, PASAD~\cite{aoudi2018truth}, and Seq2Seq-NN~\cite{kim2019anomaly}) at hand, which were all %
evaluated on the SWaT dataset~\cite{goh2016dataset}.
With \ac{IPAL}, we can apply these to two %
new datasets (WADI~\cite{ahmed2017wadi} and HAI~\cite{shin2020hai}). %
\figref{fig:casestudies:comparison} depicts the alerts of all three approaches on these three datasets and Appx.~\ref{sec:appendix:comparison} summarizes further in-depth results.

\textbf{SWaT.}
We compare the \acp{IIDS} on the first stage of the SWaT dataset, on which they were initially evaluated.
Here, we see significant differences between the operation of each \ac{IIDS}.
TABOR detects $9$ attacks, most of which are captured by Seq2Seq-NN too.
Seq2Seq-NN detects $10$ attacks while still having $17$ false alarms.
PASAD can detect $6$ scenarios, some of which were neither detected by TABOR nor by Seq2Seq-NN.
Besides detection performance, we also notice differences in how these \acp{IIDS} emit alerts.
PASAD buffers recent states internally, and Seq2Seq-NN accumulates deviations over time, delaying alerts for both approaches.
Contrary, TABOR has great precision \wrt the attack time, yet alarms are announced retrospectively due to internal segmentation.
Such differentiation only show in an in-depth side-by-side analysis as enabled by \ac{IPAL}.

\textbf{WADI.}
While the WADI dataset~\cite{ahmed2017wadi} is not unknown %
(cf.~\tabref{tab:iids-summary}), none of the three approaches evaluated it.
WADI is similar to SWaT, and we can show how simple transferring these \ac{IIDS} to a similar domain with \ac{IPAL} is again considering the first process stage of WADI only.
Here, TABOR detects $4$ scenarios, detects PASAD only $2$ attacks and Seq2Seq-NN $8$ scenarios.
During the training of PASAD, computational limits were reached due to the larger dataset size, which might contribute to the lower detection quality.
Still, the initial results of transferring an \ac{IIDS} to a new dataset with \ac{IPAL} are promising without the need to design new \acp{IIDS} from scratch.

\begin{figure*}[h!t]
	\centering
	\includegraphics[width=\textwidth]{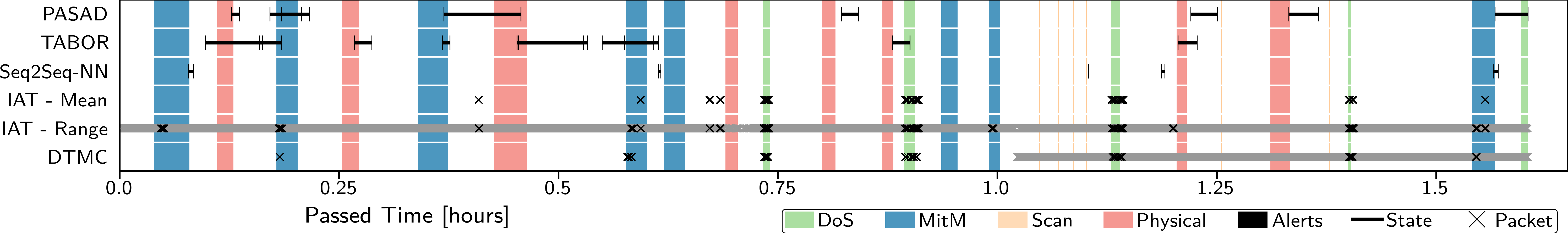}
	
	\caption{While until now considered as separate research streams, \ac{IPAL} can pool the detection capabilities of communication-based (IAT \& DTMC) and process state-aware \acp{IDS} (PASAD, TABOR, and Seq2Seq-NN), as showcased for the WDT dataset~\cite{faramondi2021hardware}.}
	\label{fig:casestudy:combination}
\end{figure*}

\textbf{HAI.}
After transferring these \acp{IIDS} to a similar domain, we now study how they generalize to other industrial settings.
Thus, we consider the fairly new HAI dataset~\cite{shin2020hai} based on a hardware-in-the-loop simulator.
As some stages of HAI are of a similar domain as SWaT and WADI, we consider the remaining boiler process used for power generation.
This dataset is much noisier than SWaT or WADI and less seldomly exhibits clear repetitive patterns challenging all \acp{IIDS}. %
TABOR does not generalize well, as it flags most parts of the dataset as an anomaly, while Seq2Seq-NN detects $2$ scenarios with $5$ false alarms.
PASAD also detects $2$ scenarios but with far more false positives ($22$).
Transferring these \acp{IIDS} to HAI with \ac{IPAL} reveals their unexplored potential to generalize to new scenarios.

\ac{IPAL} enables a detailed look at how different \acp{IIDS} compare against each other.
We even observed how differently \acp{IIDS} emit alerts motivating the use of well-chosen metrics~\cite{hwang2019time}.
Besides this in-depth comparison, \ac{IPAL} allows to apply \acp{IIDS} to new datasets quickly, and we showed that Seq2Seq-NN indeed generalizes. %

\subsection{Cooperative Communication-based and Process state-aware Intrusion Detection}
\label{sec:casestudies:combination}

As a final showcase, we explore \ac{IPAL}'s capability to compare com\-munication-based and process state-aware intrusion detection systems.
Currently, both research branches are largely disconnected (cf. \secref{sec:iids-survey}), and, as shown in \secref{sec:casestudies:comm}, a single \ac{IIDS} may not perform optimally across all attack types, motivating to combine their potentials %
even within one industrial scenario.
For our study, we train two communication-based \acp{IIDS} (IAT~\cite{lin2017timing} \& DTMC~\cite{ferling2018intrusion}) and three process state-aware \acp{IIDS} (TABOR~\cite{lin2018tabor}, Seq2Seq-NN~\cite{kim2019anomaly} \& PASAD~\cite{aoudi2018truth}) on the water distribution testbed (WDT) dataset~\cite{faramondi2021hardware}.
What sets it apart from other datasets is that it comprises of network attacks (DoS flooding, MitM to manipulate packet data, and TCP scanning), as well as, physical attacks (\eg pump breakdowns) and is thus well-suited to analyze and compare both \ac{IIDS} types.

We visualize the results of applying the five \acp{IIDS} to the WDT dataset in \figref{fig:casestudy:combination} (Appx.~\ref{sec:appendix:combination} provides further in-depth metrics).
Similar to \figref{fig:casestudies:comparison}, alerts raised by process state-aware \acp{IIDS} are visualized as time ranges.
In contrast, individual packets flagged as malicious by communication-based \acp{IIDS} are marked with $\times$.
We highlight the attack types in the dataset with different colors.

Without further tweaks, IAT range and \ac{DTMC} do not produce reliable results, as indicated by the continuous emitting of alerts (in grey).
Upon closer manual inspection, these "false" alerts can be traced back to either a missing communication sequence in the training data for DTMC or a missing margin of error for a lower-bound check in the IAT range model.
Since these alerts (in grey) are easily circumventable, we excluded them from the following evaluation and considered only alerts marked with black $\times$.

\textbf{Process state-aware.}
We expect process state-aware \acp{IIDS} to perform best on the physical attack type %
since they operate on aggregated state information (cf. \secref{sec:iids-survey-state}).
Indeed, three out of the seven attacks detected by PASAD and five out of nine attacks detected by TABOR are of the physical type.
Interestingly, Seq2Seq-NN does not detect a single physical attack even though it should be capable of doing so as it analyses the physical state.
Meanwhile, network attacks are covered insufficiently by process state-aware \ac{IIDS}, as only attacks directly aiming at physical consequences, such as some \ac{MitM} and few \ac{DoS} attacks, are detected.
In total, TABOR and PASAD detect three \ac{MitM} attacks each. %

\textbf{Communication-based.}
Like process state-aware \acp{IIDS}, com\-munication-based \acp{IIDS} detect \ac{MitM} attacks, despite not having an understanding of the underlying physical process.
DTMC detects three \ac{MitM} instances, IAT-Range five attacks, and IAT-Mean only two.
Another observation is that these \acp{IIDS} detect attacks early, likely even shortly before the attack has a real physical impact (which is detected by process state-aware \acp{IIDS}).
Regarding \ac{DoS}, all except one instance are detected by IAT and DTMC. %
Scanning, since typically considered out of scope %
for semantic \acp{IIDS}
(cf.~\secref{sec:iids:background} and \secref{sec:iids-survey}), is, as expected, hardly detectable by any approach.

Both research branches benefit from each other through \ac{IPAL}. %
Overall, process state-aware and communication-based~\acp{IIDS} detect overlapping and contradicting flavors of attack types.
Thus, they complement each other nicely, either to widen the covered attack types or to realize redundancy for equivalent attacks -- better than a single research branch can achieve on its own.
Here, \ac{IPAL} not only facilitates the generalization %
to another dataset but also, for the first time, enables the interworking of contrary research streams.

\section{Further Related Work}
\label{sec:rw}

Besides research on \acp{IIDS} covered extensively in \secref{sec:iids-survey}, our work draws inspiration from different streams of related research.
Existing tools from traditional network monitoring, such as NetFlow~\cite{RFC3954}, lack essential features and are thus not powerful enough for \acp{IIDS}.
Furthermore, adapting traditional, rule-based \ac{IDS} approaches~\cite{sommer2015spicy,roesch1999snort,Zeek} to industrial domains~\cite{yang2016multidimensional,ghaeini2016hamids,cheung2007using,fovino2010modbus,carcano2009state, gao2014cyber} is also closely related to our work.
These rule-based \acp{IDS}, however, only detect traditional (known) attacks (\cf~\secref{sec:iids:background}) and thus only complement process-aware \acp{IIDS} that are able to detect even (stealthy) attacks exploiting physical processes to cause harm~\cite{2016_urbina_limiting,giraldo2018survey}.

Furthermore, past analysis of \acp{IIDS}~\cite{giraldo2018survey, 2020_olowononi_resilient, 2020_ahmed_challenges, 2016_urbina_limiting},
partially with a focus on specific domains~\cite{ding2018survey,kaouk2019review,hu2018survey, loukas2019taxonomy, ramotsoela2018survey}, already describe problems such as data heterogeneity and evaluation bias in \ac{IIDS} research~(\cf~\secref{sec:iids:landscape}).
To analyze the ability of existing \acp{IIDS} to detect unseen sensor spoofing attacks, the only reproducibility study (considering only model-free state process-based \acp{IIDS}) known to us~\cite{erba2020no} shows significant differences between claimed generalizability and reality.
These works, in addition to reports on issues with widely-used datasets~\cite{turrin2020statistical}, motivated us to quantify these problems %
and ultimately mitigate them through protocol independence.

The problem of protocol heterogeneity in industrial communication is not exclusive to intrusion detection.
Proposals to address this issue (\eg PLC4X~\cite{PLC4X} and StreamPipes~\cite{zehnder2020streampipes, wiener2020managing}) do, however, only extract insufficient information for \acp{IIDS} and consume valuable bandwidth through polling, a limited good in many industrial scenarios~\cite{2020_armknecht_promacs, wagner2022take}.
Other work in this context is concerned with the interoperability of devices~\cite{2016_zhang_iot, 2018_viacheslav_heterogeneous, 2019_kim_microgrid, 2021_gonzalez_innovative} and proposes %
translation between different data representations.
We do, however, expect these related fields to benefit from \ac{IPAL} in the future too, \eg to reduce development costs and improve protocol support.

Lastly, Ry\v{s}av\'{y} \etal~\cite{ryvsavy2021network} propose a library to ease industrial network data preprocessing for intrusion detection.
Since their approach requires \ac{IIDS} developers to implement custom data extractors for each protocol, this solution does not tackle protocol-independence, nor does it facilitate generalizability.
Still, to motivate the library, the authors similarly acknowledge the problem of heterogeneity within the \ac{IIDS} research landscape~\cite{ryvsavy2021network} and likewise strive to move towards an interconnected \ac{IIDS} community.

\section{Conclusion}

A growing number of cyber attacks against industrial networks not only inflict substantial financial and environmental damage but even put human lives at risk~\cite{osti_1505628}.
Intrusion detection is regarded as a promising complementary protective measure to timely detect such attacks,
which is especially well suited for industrial settings due to their repetitive processes and predictable network behavior.
Consequently, a large research community gathered around industrial intrusion detection.
The produced research advancements are, however, surprisingly scattered, as many solutions are proposed for distinct communication protocols in specific industrial domains, thus hindering their transfer %
to other industrial domains.

To better understand this phenomenon, we survey $53$ scientific \acp{IIDS} and identify an unexplored potential: %
While practically operating in protocol and domain-dependent silos, theoretically, neither do \acp{IIDS} operate on the information of certain communication protocols nor are their fundamental detection methodologies specific to individual domains.
Consequently, there is a huge potential for \emph{protocol-independent} \acp{IIDS} to protect industrial networks across scenarios, not only to improve the research landscape but also to contribute %
towards widespread real-world deployments.

To unleash this potential, we propose \ac{IPAL}, our industrial protocol abstraction layer that decouples intrusion detection from domain-specific industrial communication protocols.
To this end, \ac{IPAL} transcribes %
protocol features relevant for intrusion detection into a common abstract representation. %
To show the applicability and correctness of our approach, we conducted a reproducibility study of eight \acp{IIDS} from related work, proving that \acp{IIDS} can indeed be implemented on top of \ac{IPAL}. %
Finally, with the ability to transfer \acp{IIDS} seamlessly across different industrial protocols and domains, we studied how existing \acp{IIDS} generalize to new scenarios.
We find that existing works are
\textit{(i)} indeed not confined to specific domains or protocols, but also that
\textit{(ii)} the type of attacks a given system can detect is not sufficiently studied, thus \textit{(iii)} motivating the idea to cooperatively combine approaches from previously disjoint detection domains.
With our work, we lay the foundation to break up protocol-dependence of \acp{IIDS} research and enable further studies regarding the generalization and application of \acp{IIDS} in new research or real-world scenarios.

\begin{acks}
Funded by the Deutsche Forschungsgemeinschaft (DFG, German Research Foundation) under Germany's Excellence Strategy -- EXC-2023 Internet of Production -- 390621612.
The authors would like to thank Olav Lamberts, Stefan Lenz, Tim Nebel, Leonardo Pompe, and Sven Zemanek for their work on contributing to the IPAL transcriber and \ac{IIDS} framework.
\end{acks}


\appendix
\balance

\section{Performance Comparison of Process State-aware IIDSs}
\label{sec:appendix:comparison}

In \secref{sec:casestudies:control}, we used \ac{IPAL} to compare three different process state-aware \acp{IIDS} against each other on the SWaT dataset~\cite{goh2016dataset}.
To study their potential to generalize to new industrial domains, we applied these \acp{IIDS} to two new datasets (WADI~\cite{ahmed2017wadi} and HAI~\cite{shin2020hai}).
Our analysis in \figref{fig:casestudies:comparison} depicts the alerts of these three approaches visually, which provides a decent intuition on how these approaches work in general.
To additionally confirm our observations objectively and in further detail here, \tabref{tab:ids-comparison} summarizes different metrics for all evaluated settings from that case study.

Detected attacks counts the number of scenarios overlapping with an alarm and false alarms are all alarms not overlapping with an attack.
Penalty score accumulates for all true positives the time in seconds with regions not labeled as attack within the dataset, based on the definition by Lin \etal~\cite{lin2018tabor}.
The remaining traditional metrics (accuracy, precision, recall, and f1-score) were calculated one by one for each entry within the evaluation dataset.
While these metrics may be valuable for randomized datasets, they struggle to infer the
quality of time-aware datasets~\cite{hwang2019time}, especially since approaches such as PASAD or Seq2Seq-NN detect attacks with a slight delay.

\begin{table}[H]
	\centering\footnotesize

	\begin{tabularx}{\columnwidth}{@{\hskip 0mm}l@{\hskip 1mm} X@{\hskip 1.25mm}X@{\hskip 1.25mm}X X@{\hskip 1.25mm}X@{\hskip 1.25mm}X X@{\hskip 1.25mm}X@{\hskip 1.25mm}l}
		\toprule
		\multicolumn{1}{c}{\textbf{Dataset}} & \multicolumn{3}{c}{\textbf{SWaT~\cite{goh2016dataset}}} & \multicolumn{3}{c}{\textbf{WADI~\cite{ahmed2017wadi}}} & \multicolumn{3}{c}{\textbf{HAI~\cite{shin2020hai}}} \\
		\cmidrule(lr){2-4} \cmidrule(lr){5-7} \cmidrule(lr){8-10}
		\multicolumn{1}{c}{IIDS} & \multicolumn{1}{c}{S} & \multicolumn{1}{c}{P} & \multicolumn{1}{c}{T} & \multicolumn{1}{c}{S} & \multicolumn{1}{c}{P} & \multicolumn{1}{c}{T} & \multicolumn{1}{c}{S} & \multicolumn{1}{c}{P} & \multicolumn{1}{c}{T} \\
		\midrule
		Det. Attacks 	& 10 & 6 & 9 & 8 & 2 & 4 & 2 & 2 & 35 \\
		False Alarms 	& 17 & 18 & 0 & 6 & 3 & 0 & 5 & 22 & 23 \\
		Penalty Score 	& 149 & 9374 & 1765 & 153 & 13177 & 5792 & 0 & 11669 & 214100 \\
		Accuracy [\%]	& 87.71 & 87.59 & 88.33 & 94.29 & 81.53 & 82.31 & 98.46 & 90.33 & 26.68 \\
		Precision [\%]	& 43.25 & 49.13 & 68.71 & 51.92 & 5.14 & 30.05 & 28.93 & 0.66 & 1.18 \\
		Recall [\%]		& 4.00 & 63.90 & 7.09 & 13.99 & 12.60 & 24.94 & 4.26 & 3.79 & 60.12 \\
		F1-Score [\%]	& 7.32 & 55.55 & 12.86 & 22.04 & 7.30 & 27.26 & 7.42 & 1.12 & 2.31 \\
		\bottomrule
	\end{tabularx}
	\vspace{.5em}
	\caption{\ac{IPAL} enables the comparison of process state-aware \acp{IIDS} Seq2Seq-NN (S), PASAD (P), and TABOR (T) on the SWaT dataset and in new scenarios on WADI, and HAI.}
	\label{tab:ids-comparison}
\end{table}

\begin{table}[H]
	\centering\footnotesize

	\begin{tabular}{@{\hskip 0mm}l@{\hskip 1mm} lll lll}
		\toprule
		\multicolumn{1}{c}{\textbf{IIDS Type}} & \multicolumn{3}{c}{\textbf{Process state-aware}} & \multicolumn{3}{c}{\textbf{Communication-based}} \\
		\cmidrule(lr){2-4} \cmidrule(lr){5-7}
		\multicolumn{1}{c}{IIDS} & \multicolumn{1}{c}{S} & \multicolumn{1}{c}{P} & \multicolumn{1}{c}{T} & \multicolumn{1}{c}{IAT Mean} & \multicolumn{1}{c}{IAT Range} & \multicolumn{1}{c}{DTMC} \\
		\midrule
		Det. Attacks 	& 2 & 7 & 9 & 6 & 9 & 7 \\
		False Alarms 	& 3 & 2 & 0 & 2330 & 3390 & 59 \\
		Penalty Score	& 58 & 694 & 765 & 3.67 & 2.48 & 0.06 \\
		Accuracy [\%]	& 73.99 & 67.32 & 67.22 & 75.39 & 75.39 & 75.40 \\
		Precision [\%] 	& 19.44  & 27.49 & 30.03 & 39.25 & 42.63 & 80.58 \\
		Recall [\%]		&  0.96   & 17.96 & 22.41 & 0.06 & 0.09 & 0.01 \\
		F1-Score  [\%]	&  1.83   & 21.72 & 25.67 & 0.12 & 0.19 & 0.02 \\
		\bottomrule
	\end{tabular}
	\vspace{.5em}
	\caption{For the first time, \ac{IPAL} enables the simultaneous use of process state-aware and communication-based \acp{IIDS}. To showcase the capabilities of both research branches, we trained representative \acp{IIDS} %
		study on the WDT dataset~\cite{faramondi2021hardware}.}
	\label{tab:ids-combination}
\end{table}

\section{Cooperative Communication-based and Process state-aware \acp{IIDS}}
\label{sec:appendix:combination}

In \secref{sec:casestudies:combination}, we used \ac{IPAL} to compare three process state-aware \acp{IIDS} (TABOR~\cite{lin2018tabor}, Seq2Seq-NN~\cite{kim2019anomaly} \& PASAD~\cite{aoudi2018truth}) against the complementary research branch of communication-based \acp{IIDS} represented by the IAT~\cite{lin2017timing} and DTMC~\cite{ferling2018intrusion} approaches.
To study which type of alerts these two \ac{IIDS} directions detect best, we applied them to the WDT dataset~\cite{faramondi2021hardware}, as it consists of network and physical attacks.
As before, we provide further details by summarizing the different metrics for this case study (\cf \figref{fig:casestudy:combination}) in \tabref{tab:ids-combination}.

For a short description of the used metrics, refer to the previous Appx.~\ref{sec:appendix:comparison}.
Note that process state-aware \acp{IIDS} alert over time-ranges while communication-based \acp{IIDS} mark individual network packets.
This results in poor performance of all communication-based \acp{IIDS} \wrt recall and the F1-score since this type of \ac{IIDS} emits short alerts while the metrics favor coverage over the entire attack's range.
Again, as described in \secref{sec:casestudies:combination}, the circumventable false alerts from the IAT range and DTMC approach (cf. grey annotations in \figref{fig:casestudy:combination}) are neglected for this evaluation.

\end{document}